\documentclass[a4paper,fleqn,usenatbib]{mnras}

\usepackage[T1]{fontenc}
\usepackage{ae,aecompl}

\usepackage{graphicx}	
\usepackage{amsmath}	
\usepackage{amssymb}	
\usepackage{multirow}
\usepackage{booktabs}
\usepackage{txfonts}

\newcommand{\mjup}{\mathrm{M_{Jup}}}
\newcommand{\excs}{\extracolsep{\fill}}

\title[Possible origin of exozodis]{Inner mean-motion resonances with eccentric planets: \\ A possible origin for exozodiacal dust clouds}

\author[V. Faramaz et al.]{
 V. Faramaz,$^1$\thanks{E-mail: virginie.faramaz@gmail.com}
 S. Ertel,$^2$
 M. Booth,$^{1,3}$
 J. Cuadra,$^1$
 C. Simmonds$^1$,
\\
$^1$ Instituto de Astrof\'isica, Pontificia Universidad Cat\'olica de Chile, Vicu\~na Mackenna 4860, 7820436 Macul, Santiago, Chile\\
$^2$ Steward Observatory, Department of Astronomy, University of Arizona, 933 N. Cherry Ave, Tucson, AZ 85721, USA\\
$^3$ Astrophysikalisches Institut und Universit\"atssternwarte, Friedrich-Schiller-Universit\"at Jena, Schillerg\"a{\ss}chen 2-3, \\
07745 Jena, Germany
}

\date{Accepted 2016 November 2. Received 2016 November 2; in original form 2016 July 25}

\pubyear{2016}

\begin{document}
\label{firstpage}
\pagerange{\pageref{firstpage}--\pageref{lastpage}}
\maketitle

\begin{abstract}
High levels of dust have been detected in the immediate vicinity of many stars, both young and old. A promising scenario to explain the presence of this short-lived dust is that these analogues to the Zodiacal cloud (or exozodis) are refilled in situ through cometary activity and sublimation. As the reservoir of comets is not expected to be replenished, the presence of these exozodis in old systems has yet to be adequately explained. 
It was recently suggested that mean-motion resonances (MMR) with exterior planets on moderately eccentric ($\mathrm{e_p}\gtrsim 0.1$) orbits could scatter planetesimals on to cometary orbits with delays of the order of several 100 Myr. Theoretically, this mechanism is also expected to sustain continuous production of active comets once it has started, potentially over Gyr-timescales.\\
We aim here to investigate the ability of this mechanism to generate scattering on to cometary orbits compatible with the production of an exozodi on long timescales.
We combine analytical predictions and complementary numerical N-body simulations to study its characteristics.\\
We show, using order of magnitude estimates, that via this mechanism, low mass discs comparable to the Kuiper Belt could sustain comet scattering at rates compatible with the presence of the exozodis which are detected around Solar-type stars, and on Gyr timescales. We also find that the levels of dust detected around Vega could be sustained via our proposed mechanism if an eccentric Jupiter-like planet were present exterior to the system's cold debris disc. 
\end{abstract}




\section{Introduction}

The zodiacal cloud is the name given to the population of small dust grains that permeate the Solar System's terrestrial planet region. The source of these grains has long been debated, with the asteroid belt, Kuiper belt, comets and interplanetary grains all being suggested as possible sources. Recent work by \citet{2010ApJ...713..816N} and \citet{2013MNRAS.429.2894R} shows that the majority of the dust (70-95\%) comes from Jupiter Family Comets.

Although often related to dust in the habitable zone, our zodiacal dust is distributed over a wide range of radii from the sun. It ranges from the Asteroid belt inwards, through the habitable zone, to the dust sublimation radius at $\sim4$ Solar radii. The radial dust distribution globally follows an inward slowly increasing power-law that is locally modified by the interaction with the inner, rocky planets and the local production of dust through comet evaporation \citep{1998ApJ...508...44K}. In the innermost regions it forms the Fraunhofer corona \citep[F-corona,][]{1998EP&S...50..493K}.

In analogy to the zodiacal light, warm and hot dust around other stars is called exozodiacal dust (exozodi, emission: exozodiacal light). To observe this faint emission close to nearby main sequence stars, high angular resolution and contrast are needed. Thus, interferometric observations are the method of choice. In the mid-infrared the emission of habitable zone dust can be detected using nulling interferometry \citep{2014ApJ...797..119M}. In the near-infrared hotter dust closer to the star can be detected using optical long baseline interferometry \citep[e.g.,][]{2006A&A...452..237A}. The largest statistical survey for exozodis so far has been carried out in the near-infrared where a detection rate of ~15 - 20\% has been found \citep{2013A&A...555A.104A,2014A&A...570A.128E}.

For A stars, there is no apparent correlation between the presence of an exozodi and the age of the system and for FGK stars there is tentative evidence of a trend for older systems to be more likely to show evidence for an exozodi  \citep{2014A&A...570A.128E}. This is particularly surprising as mid-infrared observations show a clear decline in presence of excess with age \citep{2006ApJ...653..675S,2009ApJS..181..197C,2009ApJ...697.1578G} as is expected from a population evolving through collisions \citep{2007ApJ...658..569W,2008ApJ...673.1123L}. If, as in the Solar System, comets are also responsible for the dust in exozodis then this appears to contradict with what would be expected, since the reservoir of planetesimals is not replenished, and the dust feeding rates should eventually diminish with time.

This has motivated several recent studies to investigate the origin of these exozodis, with particular focus on how they can be present in systems with ages of order of 100 Myr - 1 Gyr. A possible explanation is that the production of comets is delayed on large timescales (several 100 Myr).
This can be achieved if the system is in a phase similar to that of the Solar System Late Heavy Bombardment (LHB). In the Nice model of \citet{2005Natur.435..466G}, the Solar system was originally more compact, with Neptune orbiting inside Uranus, and suffered major changes at $\sim 800\,$Myr. This lead Neptune to jump beyond Uranus, but also lead the four giant planets to adopt more eccentric orbits, thus enhancing close encounters with planetesimals and generating a short phase of intense production of active comets. However, whilst this mechanism appears to be a good candidate to explain the presence of exozodis in some old systems where there are other reasons to believe that a LHB-like event took place such as in the $\eta\,$Crv system \citep{2009MNRAS.399..385B,2012ApJ...747...93L}, the detection rate of exozodis would be only $\sim 0.1\%$ if all were produced due to LHB-like events \citep{2013MNRAS.433.2938B}.

Another possibility is that the levels of dust are maintained throughout the lifetime of these systems, as suggested by \citet{2012A&A...548A.104B}, who investigated scattering of planetesimals from an outer cold belt to the inner parts of a system by a chain of planets.It was found that the necessary scattering rates could be obtained, although they required contrived multi-planet system architectures involving chains of tightly packed low-mass planets. The rates were found to be higher, or sustained on longer timescales if the outermost planet of the chain happens to migrate outwards into the belt while scattering planetesimals \citep{2014MNRAS.441.2380B,2014MNRAS.442L..18R}. However, this requires additional conditions for migration to take place which depend highly on the mass of the planet and the disc surface density.

In \citet{2015A&A...573A..87F}, a dynamical process is presented that shows that inner mean-motion resonances (MMR) with a planet on an orbit at least moderately eccentric ($\mathrm{e_p}\gtrsim 0.1$), is a valuable route to set planetesimals on highly eccentric orbits, potentially cometary-like, with periastrons small enough for bodies to sublimate. This mechanism combines several strengths : in contrast with scattering of the reservoir by a chain of planets, it involves only one single planet, and the generation of cometary-like orbits through this process can be delayed by timescales as large as several 100 Myr, as for a LHB-like event. In addition, it is expected to place bodies on cometary-like orbits continuously, potentially over large timescales. Therefore, this mechanism could provide a robust explanation for the presence of exozodis, especially in systems older than 100 Myr.

The general question of the ability of this mechanism to generate a flux of active comets compatible with the presence of an exozodi is addressed in this paper, with particular focus on old systems. We present this mechanism in more details in Sect.~\ref{sec:analytical}, as well as the analytical background. We make predictions on the ability of a given MMR with a perturber of eccentricity 0.1 to scatter planetesimals on cometary-like orbits. These predictions are complemented by a numerical analysis, and we present a study case of a given MMR, namely the 5:2, in Sect.~\ref{sec:results}. This will allow us to determine the efficiency of this mechanism in a quantitative manner (rates, timescales) and as a function of the characteristics of the planet (semi-major axis, mass). We compare achievable scattering rates with observations and show applications of this framework to the cases of a low mass disc around a Sun-like star and the Vega system in Sect.~\ref{sec:applis}. Finally, in Sect.~\ref{sec:conclusion}, we discuss the efficiency of other MMRs and the impact of the eccentricity of the planet, before drawing our conclusions.


\section{Analytical study}\label{sec:analytical}

In this section, we will present in more detail the mechanism of \citet{2015A&A...573A..87F}, and in particular we will show the analytical predictions that can be made on the ability of a given perturber and a given MMR to set planetesimals on cometary orbits. The question of whether this mechanism can induce sufficient feeding rates and the determination of its characteristic timescales (delays and duration) will be addressed using a numerical analysis in the next sections.

\subsection{Orbital evolution of planetesimals in MMRs with an outer eccentric planet}

MMRs between a planetesimal and a planet, usually noted $\mathrm{n:p}$, where $\mathrm{n}$ and $\mathrm{p}$ are integers, concern bodies with orbital periods achieving the $\mathrm{p/n}$ commensurability with that of the planet. Therefore, MMRs occur at specific locations relative to the orbit of the planetary perturber. The integer $\mathrm{q=|n-p|}$ is called the order of the resonance. Resonances with $\mathrm{n>p}$ correspond to \emph{inner} resonances, that is, planetesimals orbiting inside the orbit of the planet, while $\mathrm{n<p}$ denotes \emph{outer} resonances.

\begin{figure*}
\centering
\makebox[\textwidth]{\includegraphics[scale=0.5]{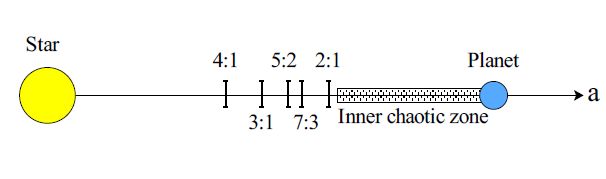}}
\caption[]{Schematic edge-on view of the systems we focus on in this study. According to Eqs.~(\ref{eq:MMRloc}) and ~(\ref{eq:chaosin2}), both the location of the MMRs and the inner edge of the chaotic zone are proportional to the semi-major axis of the planet $\mathrm{a_p}$. The size of the chaotic zone and the relative distance between its inner edge and the location of the MMRs, depend on the mass of the planet, which was set to $\mathrm{m_p}=0.5\,\mjup$, and on its eccentricity, which was set to $\mathrm{e_p}=0.1$.}
\label{fig:MMR_schema}
\end{figure*}

The process we focus on here implies planetesimals initially on almost circular orbits and protected from close-encounters with the planet, which they are usually assumed to be at the end of the protoplanetary phase because of the eccentricity damping action of the gas. If these planetesimals are in inner MMR with an eccentric planet as shown in Fig.~\ref{fig:MMR_schema}, they may then be driven to more eccentric orbits.

When a planetesimal is trapped in MMR with a planet that is on a circular orbit, the evolution of its eccentricity is coupled to that of the semi-major axis. Since this semi-major axis undergoes small amplitude librations ($\la 0.1$AU) around the exact resonance location, its eccentricity undergoes small amplitude variations as well. However, as soon as the eccentricity of the planet is larger than $\sim 0.05$, the amplitude of the variations of the eccentricity can be increased up to large values \citep[see][for details]{1995Icar..115...60M,1996Icar..120..358B}.

This can be understood from analytical considerations, and in particular by using phase-space diagrams.
These consist of lines of constant energy, and are obtained using the Hamiltonian of the system which has been reduced to a single degree of freedom beforehand. The resulting Hamiltonian thus depends only on $\mathrm{(\nu,e)}$, where $\nu$ is the longitude of periastron of the planetesimal with respect to that of the planet, and $\mathrm{e}$ is the eccentricity of the planetesimal. This means that the precession cycle of a planetesimal is coupled with its eccentricity evolution. In order to easily visualize the co-evolution of these two quantities, the Hamiltonian can then be evaluated for a grid of $(\nu,e)$, and level curves can be drawn \citep[for more details on the theoretical background, see ][]{1993CeMDA..57...99M,1995Icar..114...33M,1996Icar..120..358B,2000Icar..143..170B,2015A&A...573A..87F}.

\begin{figure}
\centering
\includegraphics[width=0.8\columnwidth]{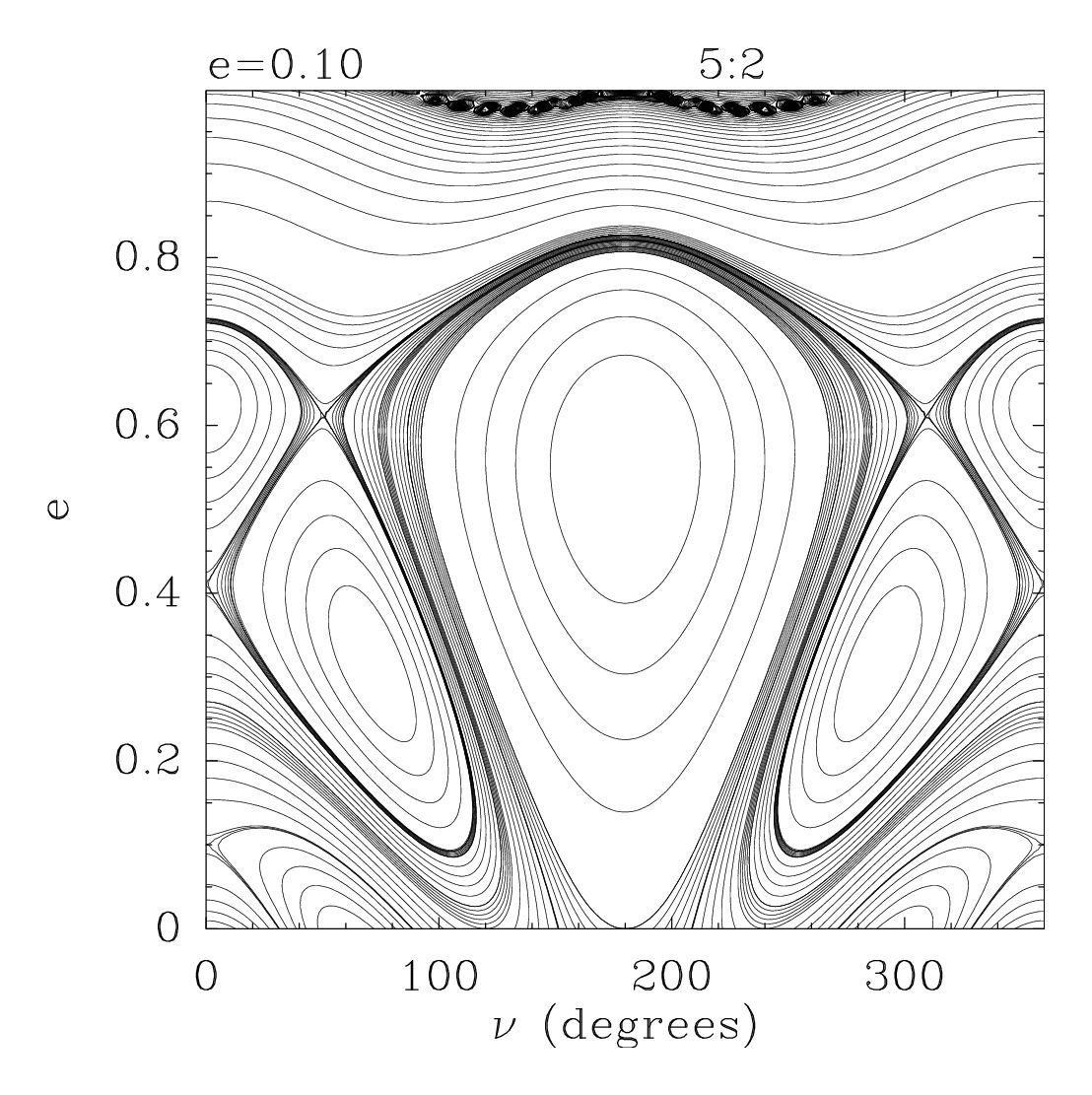}
\caption[]{Theoretical phase space diagram of the 5:2 MMR with a planet of eccentricity $\mathrm{e_p=0.1}$. This diagram is made of iso-Hamiltonian curves, so that the behaviour of the orbit of a planetesimal in 5:2 MMR with a planet that has an eccentricity of 0.1 can be evaluated by following the trajectories in the $\mathrm{(\nu,e)}$ phase-space, where $\nu$ is the longitude of periastron of the planetesimal with respect to that of the planet, and $\mathrm{e}$ is the eccentricity of the planetesimal.}
\label{fig:phase_space52}
\end{figure}

In Fig.~\ref{fig:phase_space52}, we display the trajectories of a planetesimal in the $\mathrm{(\nu,e)}$ phase-space, when trapped in 5:2 MMR with a planet of eccentricity $\mathrm{e_p}=0.1$. As can be read on the Y-axis, the eccentricity of the planetesimal can be increased from an initially quasi circular orbit up to values $\sim 0.85$. This occurs when the orbit of the planetesimal is anti-aligned with that of the planet, that is, when there is an angle of $\nu=180^{\circ}$ between the periastron of the planetesimal and that of the planet (X-axis).
Since the semi-major axis of the planetesimal only undergoes small amplitude librations, it is roughly constant. Therefore, when the eccentricity of the planetesimal increases, its periastron diminishes, while its apastron increases.
Note that these theoretical paths in the $(\mathrm{\nu,e})$ phase-space are derived making the assumption that the amplitude of libration of the resonance is zero, so that these paths are ideal paths compared to the real ones that particles will adopt in our numerical simulations, and which will oscillate around the ideal path, as shown in Fig. 13 of \citet{2000Icar..143..170B}.

In this paper we are interested in the rate of dust input into an exozodi. We are working on the assumption that comets provide this dust in a similar way to how comets in the Solar System provide dust to the zodiacal cloud \citep[see ][]{2010ApJ...713..816N}. It is thought that the main mass loss for comets is their spontaneous disruption rather than the continual loss of dust through activity \citep[e.g., ][]{1980A&A....85..191W,1994Icar..108..265C,2004come.book..301B,2005ASSL..328.....F,2006mspc.book.....J}. Nonetheless, the most likely causes of disruption are related to the physical changes a comet undergoes whilst it is active \citep{1980A&A....85..191W,2004come.book..301B} and so we will focus on determining the number of planetesimals that become active. A comet becomes active when it comes close enough to the star for water ice to sublimate, which means that its periastron has to decrease to less than $\mathrm{q_{sub}}$. This distance varies with the luminosity of the star. For Solar-type stars, for example, comets are expected to undergo sublimation at separations smaller than 3 AU \citep[see e.g., ][]{2016arXiv160403790M}. If the eccentricity increase is sufficient, there will be two possible mechanisms for a planetesimal to develop onto an orbit that results in it becoming an active comet.

Firstly, a MMR can potentially drive the periastron of a planetesimal to values smaller than $\mathrm{q_{sub}}$, which is a direct route to set planetesimals on to cometary orbits. Secondly, a MMR can also potentially drive planetesimals' apastrons to values such that their orbit crosses the chaotic zone of the planet. The chaotic zone is the region around the planet where MMRs overlap. Interactions occurring there between a planet and a planetesimal can lead to major changes in the orbit of the planetesimal, that is, it can be scattered out of MMR on to a potentially comet-like orbit. This is an indirect mechanism in two steps, with a first phase of eccentricity increase while in MMR, followed by a scattering event.

In either case, planetesimals set on cometary orbits will endure cometary evaporation during phases of close approach to their host star, deposit dust, maybe even disrupt, and contribute to an exozodi. We describe hereafter the direct and indirect placement of planetesimals on to cometary orbits.

\subsection{Direct placement on cometary orbits}

Using Kepler's third law, the theoretical semi-major axis $\mathrm{a_{MMR}}$ of a given $\mathrm{n:p}$ MMR is linked to the planet semi-major axis by :

\begin{equation}\label{eq:MMRloc}
\mathrm{a_{MMR}}= \mathrm{a_p} \mathrm{\left(\frac{p}{n} \right)^{2/3}}\qquad.
\end{equation}

\begin{figure*}
\centering
\makebox[\textwidth]{\includegraphics[width=0.4\textwidth,height=0.4\textwidth]{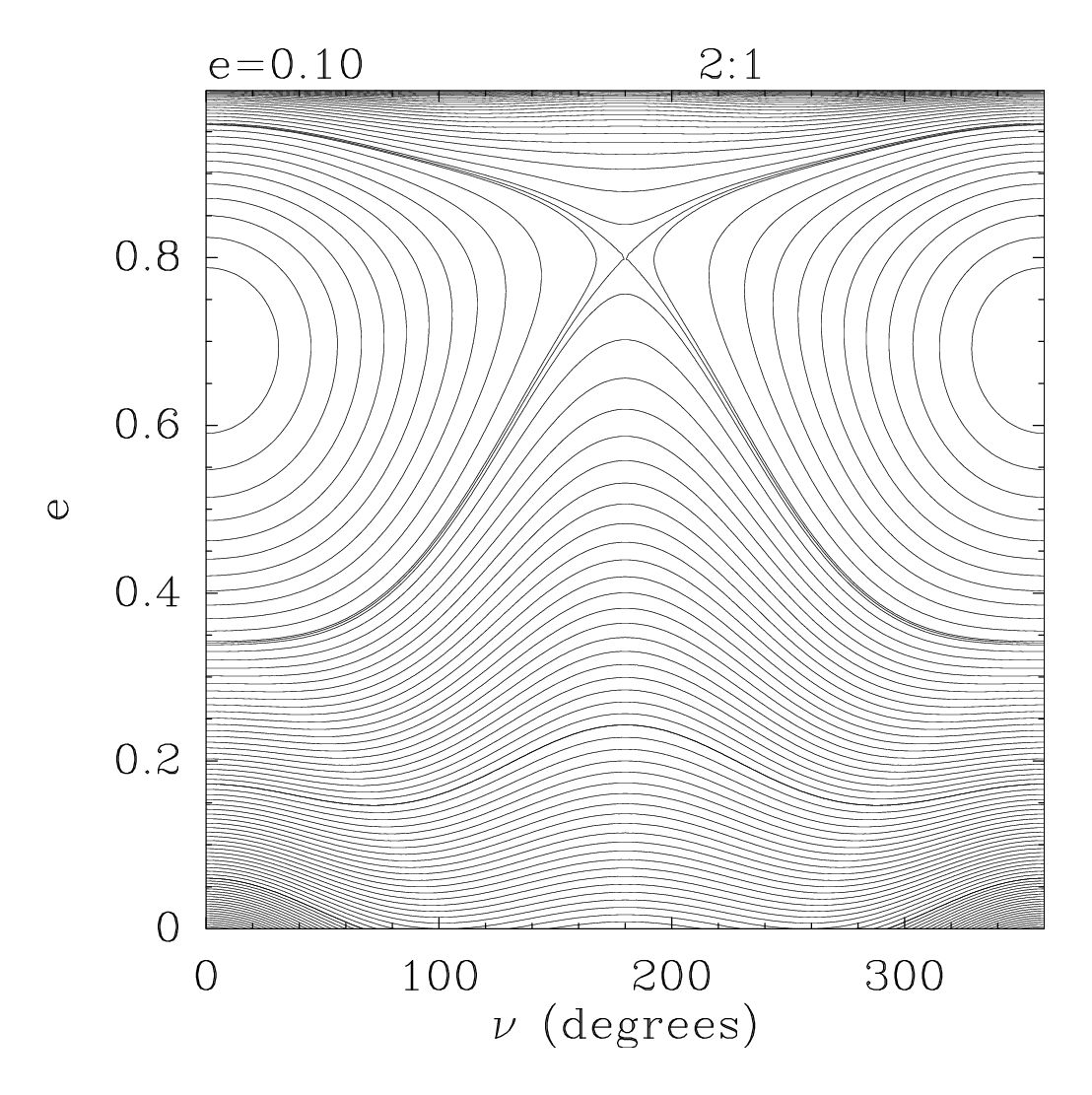}
\includegraphics[width=0.4\textwidth,height=0.4\textwidth]{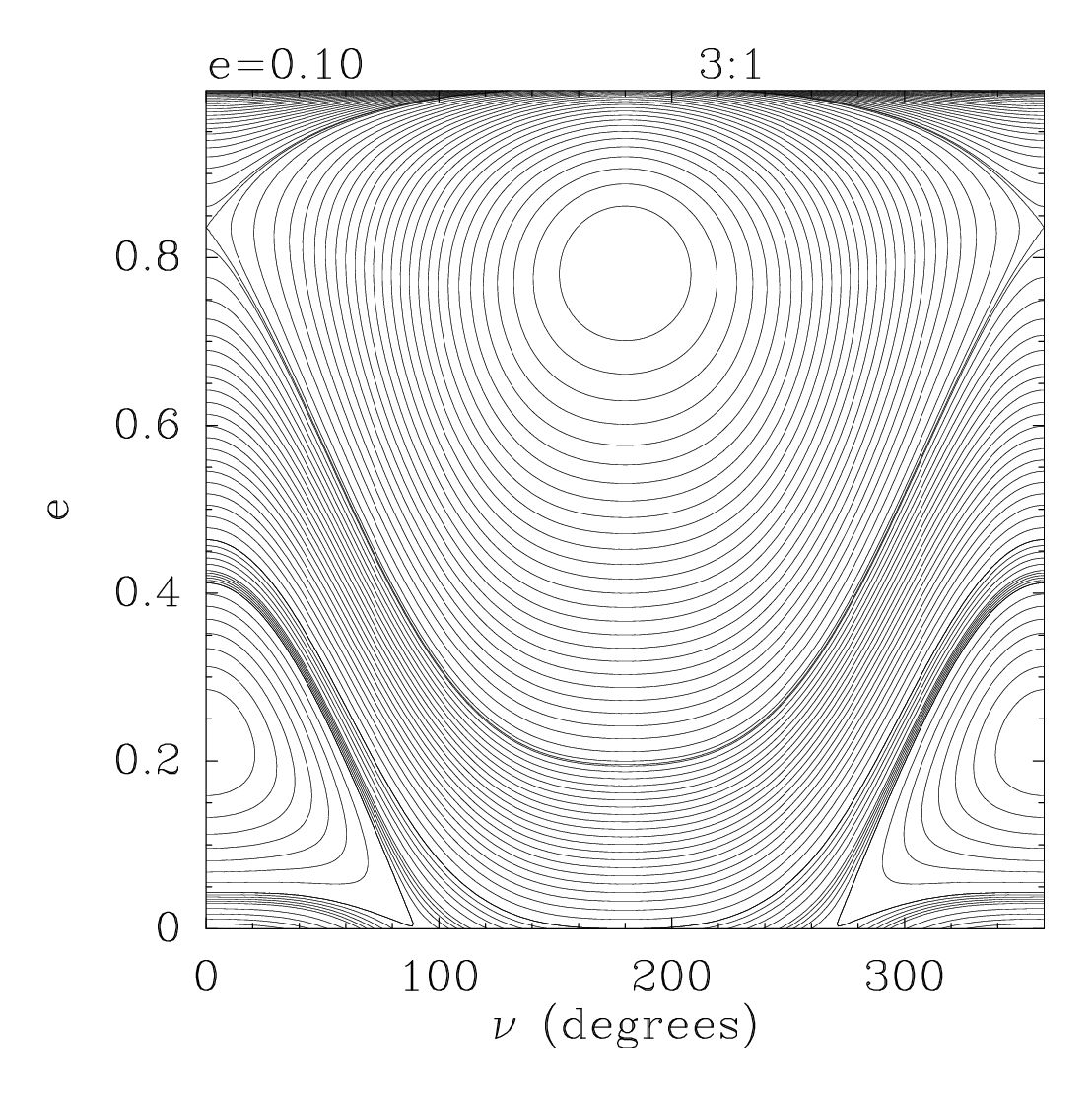}
\includegraphics[width=0.4\textwidth,height=0.4\textwidth]{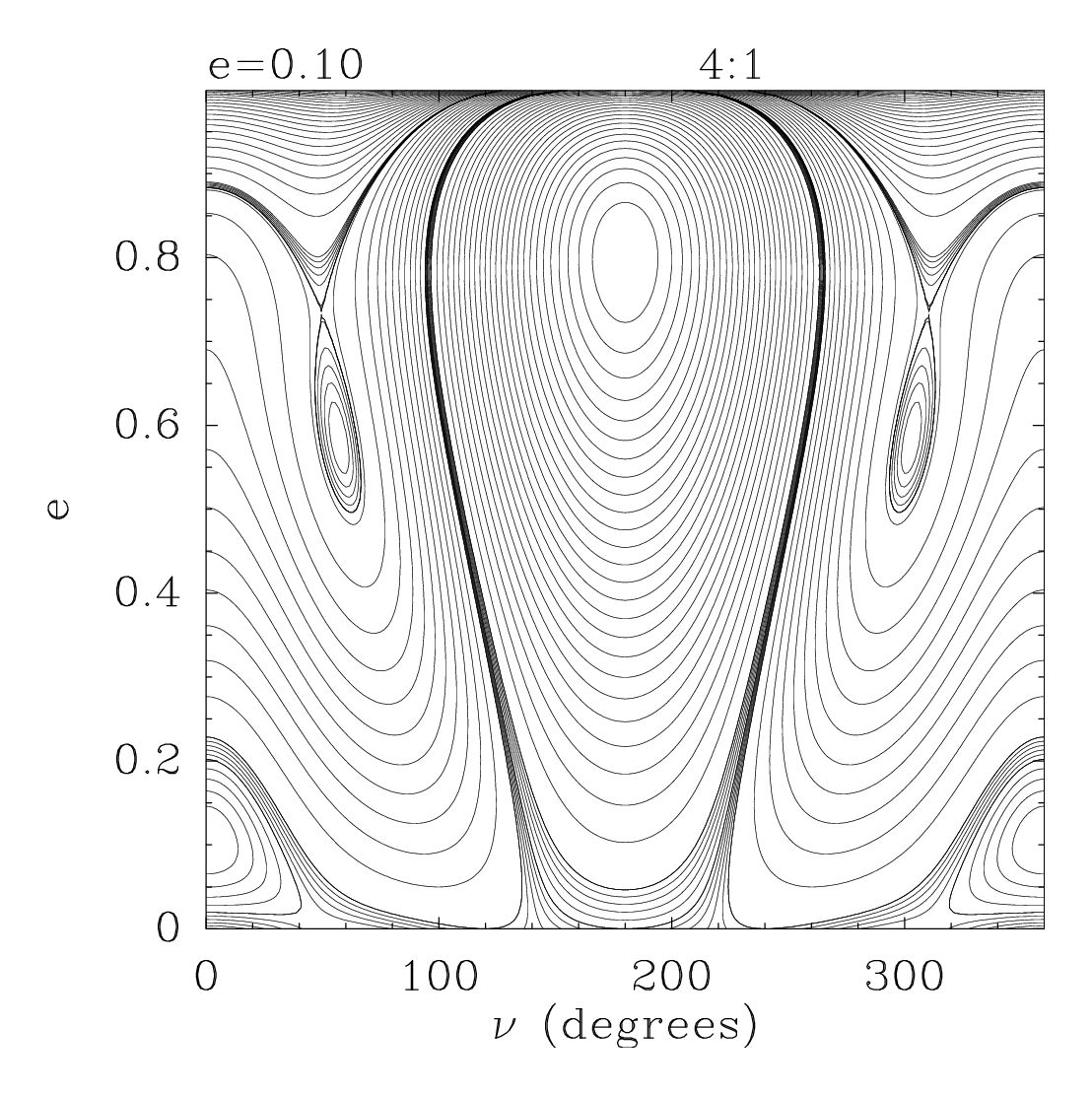}}
\makebox[\textwidth]{
\includegraphics[width=0.4\textwidth,height=0.4\textwidth]{MMR52.jpg}
\includegraphics[width=0.4\textwidth,height=0.4\textwidth]{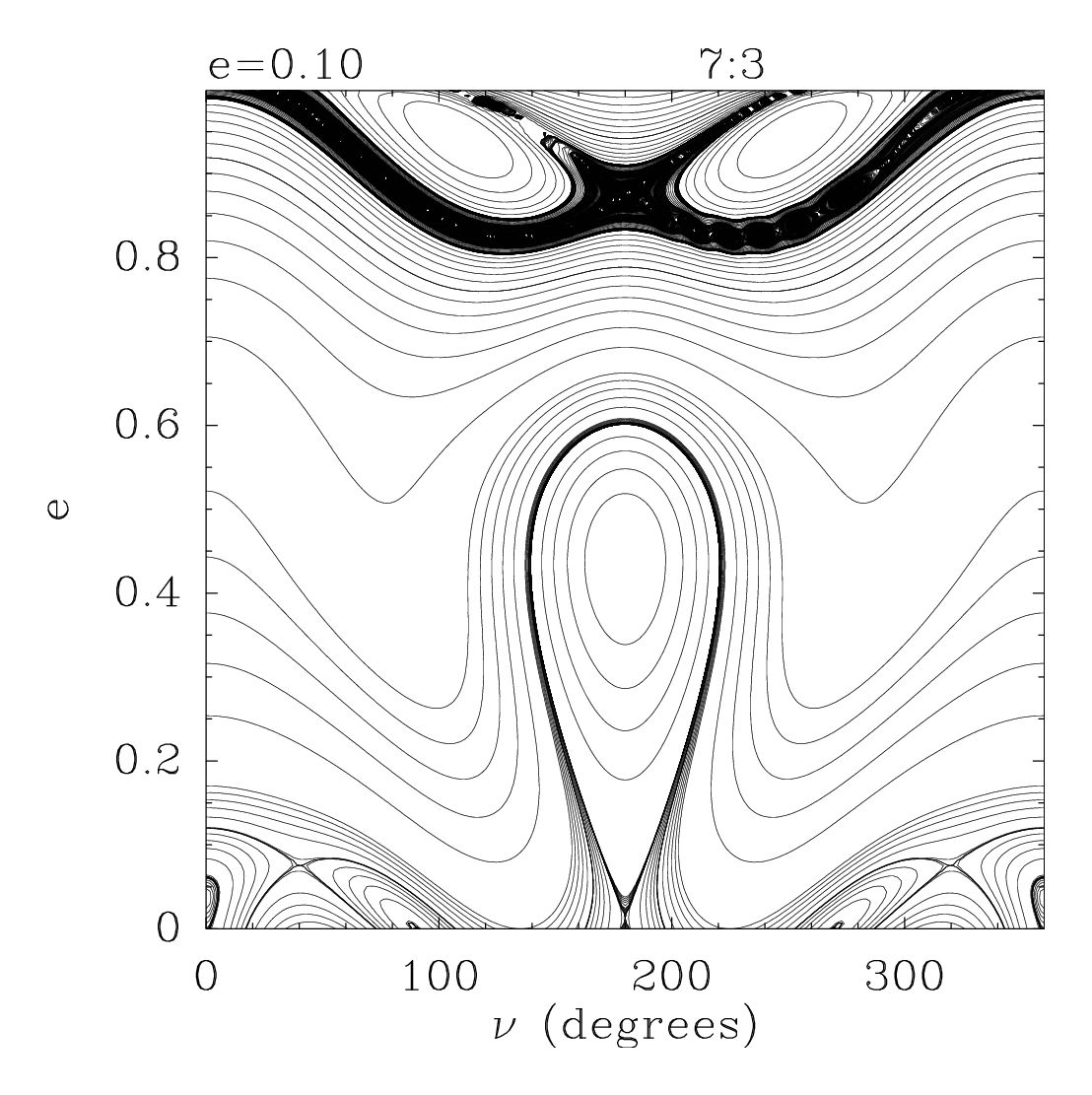}}
\caption[]{Theoretical phase-space diagrams (see caption to Fig.~\ref{fig:phase_space52} of diverse MMRs with a planet of eccentricity $\mathrm{e_p}=0.1$.}
\label{fig:phase_space1}
\end{figure*}

For a given MMR, one can estimate from the phase space diagram the maximum eccentricity $\mathrm{e_{max}}$ that a planetesimal can acquire when starting from a low eccentricity orbit $(\mathrm{e} \lesssim 0.05)$. In Fig~\ref{fig:phase_space1}, we compare phase-space diagrams of different MMRs, namely those shown on Fig.~\ref{fig:MMR_schema}. One can see that the MMRs which induce the largest eccentricities are the 4:1, 3:1, 7:3 and 5:2 MMRs.
One can then retrieve the minimum periastron $\mathrm{q_{min}=a_{MMR}(1-e_{max})}$ that can be achieved.
The 4:1 MMR is the only one that is able to increase the eccentricity of a planetesimal up to values $\sim 1$. This means that this resonance can produce cometary orbits directly no matter what the semi-major axis of the planet is, without the need for an additional scattering event, as is necessary for the mechanism we consider in the rest of this paper. The other MMRs can do this too, but for a very limited range of planet semi-major axes. For a Solar-type star, where $\mathrm{q_{sub}}=3\,$AU, $\mathrm{q_{min}}<\mathrm{q_{sub}}$ when $\mathrm{a_p} \lesssim 15\,$AU for the 3:1 and 7:3 MMRs, and $\mathrm{a_p} \lesssim 35\,$AU for the 5:2 MMR. The 2:1 MMR does not trigger larger eccentricities $(\sim 0.1)$, hence it would directly induce the production of active comets for semi-major axes smaller than $\sim 5\,$AU.

In the case of direct cometary production, all the planetesimals in MMR that can be driven on to cometary orbits will become comets during their first precession cycle. Therefore, one will expect a sharp peaked profile of the time-dependent cometary event rate, and an activity starting very quickly after the planet starts acting upon the planetesimals. \citet{1996Icar..120..358B} found that the timescale for a planetesimal to reach its maximum eccentricity was of the order of $10^3-10^4$ planetary orbital periods, and depends also on the MMR itself. This translates into timescales of the order of 1 Myr at most for planets of semi-major axis smaller than 10-25 AU.
In this case, the timescales on which production of active comets will be sustained may be incompatible with the presence of an exozodi in old systems. 
However, this timescale is expected as well to increase with smaller planet-star mass ratios, and may also reach large values. Therefore, we will determine numerically in Sect.~\ref{sec:results} if these values are large enough for direct placement on cometary orbits to produce active comets on timescales of several 100 Myr.

\subsection{Indirect mechanism: additional scattering event and large timescales}\label{sec:direct}

In cases where $\mathrm{q_{min}}$ remains greater than $\mathrm{q_{sub}}$, a planetesimal can still be placed on to an active comet orbit if the apastron increases to the point where it crosses the chaotic zone of the planet.
If the planetesimal and the planet actually encounter in this orbital configuration, the planetesimal will be scattered by the planet, potentially on to a cometary-like orbit. 
When in MMR, a planetesimal will require time to travel through its phase-space and to reach a position where its orbit crosses the chaotic zone of the planet. The planetesimal will not necessarily encounter the planet at this point and it may take several precession cycles before a close encounter with the planet occurs. The number of cycles that is necessary will depend on the initial position of the planetesimal in the phase-space. In a realistic situation, this mechanism applies to a collection of planetesimals in MMR, which span a whole range of initial conditions. Therefore, at each precession cycle, different planetesimals will be brought within reach of the planet. It is thus expected that once the planetesimals which encounter the planet during the first precession cycle are scattered on to cometary orbits, the process generates a continuous flux of comets. The timescale for a precession cycle to complete is expected to depend on the semi-major axis of the planet, and also on the mass ratio between the planet and the star. This timescale can reach several 100 Myrs when the planet is Saturn- Neptune-sized and at large separation from the star \citep[$\sim 100\,$AU][]{2015A&A...573A..87F}. Consequently, a first feature of this mechanism is that scattering events and cometary events can start occurring very late in the history of a system. 
But more importantly, since the production of cometary-like orbits is expected to be continuous on at least several precession timescales, it could be sustained on very large, potentially Gyr timescales, in contrast to a short period of intense scattering on cometary orbits as in LHB-like events.
In addition, the cometary orbits, once produced, are expected to have a potentially significant dynamical lifetime against ejection \citep[several 0.1-10 Myr][]{2015A&A...573A..87F}, and the planetesimals are not lost onto the star but continuously deposit dust at each passage. However, their lifetime against evaporation may differ from the dynamical lifetime of the orbit. These aspects will be discussed in Sect.~\ref{sec:applis}. 

This indirect mechanism that we wish to study here applies to planets at large separations from their host star (several tens of AU).
There are usually two possible explanations for the presence of a planet at large distance from its host star, that is, with semi-major axis $\mathrm{a_p}$ of several tens of AU, or even greater than a hundred AU. These are outward migration and planet-planet scattering. However, migrating planets tend to have nearly circular orbits \citep{1997Icar..126..261W,2003ApJ...588..494M,2008ApJ...673..487I,2009ApJ...705L.148C}. 
Therefore, planet-planet scattering is in general favoured over the migration scenario, because it could explain the large orbital distance as well as the eccentricity of the perturber \citep[see for instance][]{1996Sci...274..954R,1997ApJ...477..781L,2008ApJ...686..621F,2008ApJ...686..580C,2008ApJ...686..603J}.

One can expect a major instability such as one triggering planet-planet scattering to severely damage discs. If the planet-planet scattering is too violent then it would likely scatter all of the material from the disc and wipe it out. However, dynamical simulations of planetary systems do show a range of outcomes including many cases with moderate level of scattering that only partially depletes the disc \citep{2011A&A...530A..62R}. We will therefore place our work under the assumption that disc material remains available after the instability for our mechanism to take place.
Note as well that the Solar System is a system where scattering has likely taken place and depleted the comet reservoir \citep{2005Natur.435..466G,2009MNRAS.399..385B} and we will show in Section~\ref{sec:comp_obs} that such a faint disc could still provide enough material for a detectable exozodi.

\subsection{Indirect mechanism: analytical predictions}

The ability of a given planet to scatter planetesimals originating from a given MMR, potentially on a cometary-like orbit, can be predicted by determining whether the apastron of planetesimals in MMR can be increased up to the point where it crosses the chaotic zone of the planet.

Using the definition $\mathrm{Q=a_{MMR}(1+e)}$ of the apastron of a planetesimal and combining with $\mathrm{e_{max}}$ as defined above and Eq.~\ref{eq:MMRloc}, one can obtain the maximum apastron that the orbit of the planetesimal can reach :

\begin{equation}\label{eq:Qmax}
\mathrm{Q_\mathrm{max}= a_\mathrm{p} \left(\frac{p}{n} \right)^{2/3} (1+e_{max}) }\qquad.
\end{equation}

On the other hand, following \citet{2014MNRAS.443.2541P}, the inner chaotic zone of the planet starts at a semi-major axis $\mathrm{a_{chaos,in}}=\mathrm{q_p-\Delta a_{chaos,in}}$, where, in the case of an eccentric planet, $\mathrm{\Delta a_{chaos,in}}$ corresponds to several times the Hill radius at periastron of the eccentric planet, $\mathrm{R_{H,q_p}}$, which \citet{2014MNRAS.443.2541P} find to be :

\begin{equation}\label{eq:rhill_peri}
\mathrm{R_{\mathrm{H,q}} \approx a_{\mathrm{p}}(1-e_{\mathrm{p}})\left[ \frac{m_{\mathrm{p}}}{(3+e_{\mathrm{p}})M_{\star}} \right]^{\frac{1}{3}}} \qquad,
\end{equation}

and thus,

\begin{equation}\label{eq:chaosin1}
\mathrm{a_{chaos,in}=q_p-k\times R_{\mathrm{H,q}}} \qquad,
\end{equation}

where the factor $\mathrm{k}$ has to be determined numerically. \citet{2014MNRAS.443.2541P} focused on the size of a planet's outer chaotic zone, and found that $\mathrm{k\simeq 5}$ when setting $\mathrm{a_{chaos,out}=Q_p+k\times R_{\mathrm{H,Q}}}$, where $\mathrm{R_{\mathrm{H,Q}}}$ is the Hill radius at apastron, that shall be used in the case of the outer chaotic zone. They did not carry out the study for the inner chaotic zone. However, since we simply want to make qualitative predictions that shall be checked later numerically, we will use $\mathrm{k=5}$ in our calculations. We obtain :

\begin{equation}\label{eq:chaosin2}
\mathrm{a_{chaos,in} \approx a_{\mathrm{p}}(1-e_{\mathrm{p}})\left[1-k\left( \frac{m_{\mathrm{p}}}{(3+e_{\mathrm{p}})M_{\star}} \right)^{\frac{1}{3}}\right]} \qquad,
\end{equation}

where $\mathrm{M_{\star}}$ is the mass of the central star.

The condition that must be verified for a given MMR to lead planetesimals to cross the chaotic zone of a given planet, that is $\mathrm{Q_{max}\geq a_{chaos,in}}$, finally reads:

\begin{equation}\label{eq:ability1}
\mathrm{\left(\frac{p}{n} \right)^{2/3} (1+e_{max}) \geq (1-e_{\mathrm{p}})\left[1-k\left( \frac{m_{\mathrm{p}}}{(3+e_{\mathrm{p}})M_{\star}} \right)^{\frac{1}{3}}\right]}\qquad.
\end{equation}

This condition depends on the MMR itself, on the maximum eccentricity this MMR can induce on a planetesimal, and on the mass ratio between the planet and the star. 

Note that it does not depend on the semi-major axis of the planet. This was expected, since according to Eqs.~(\ref{eq:MMRloc}) and ~(\ref{eq:chaosin2}), all the quantities involved here, i.e. the location of the MMR and the inner edge of the chaotic zone, are proportional to the semi-major axis of the planet $\mathrm{a_p}$. The size of the chaotic zone and the relative distance between a resonance location and the inner edge of the chaotic zone, depend on the mass and eccentricity of the planet. Thus if scattering events are expected for a given combination of planetary mass and eccentricity, they will be expected for any planetary semi-major axis.

The condition stated in Eq.~\ref{eq:ability1} can be rewritten to find $\mathrm{e_{crit}}$, the minimum eccentricity a planetesimal should acquire when in a given MMR to cross the chaotic zone of the planet, as a function of the planet mass and eccentricity :

\begin{equation}\label{eq:ability2}
\mathrm{e \geq e_{crit} = \left( \frac{n}{p} \right)^{2/3}(1-e_{\mathrm{p}})  \left(  1- k\left[ \frac{m_{\mathrm{p}}}{(3+e_{\mathrm{p}})M_{\star}} \right]^{\frac{1}{3}} \right) - 1 }\qquad.
\end{equation}

\begin{figure*}
\centering
\makebox[\textwidth]{\includegraphics[width=0.5\textwidth,height=0.5\textwidth]{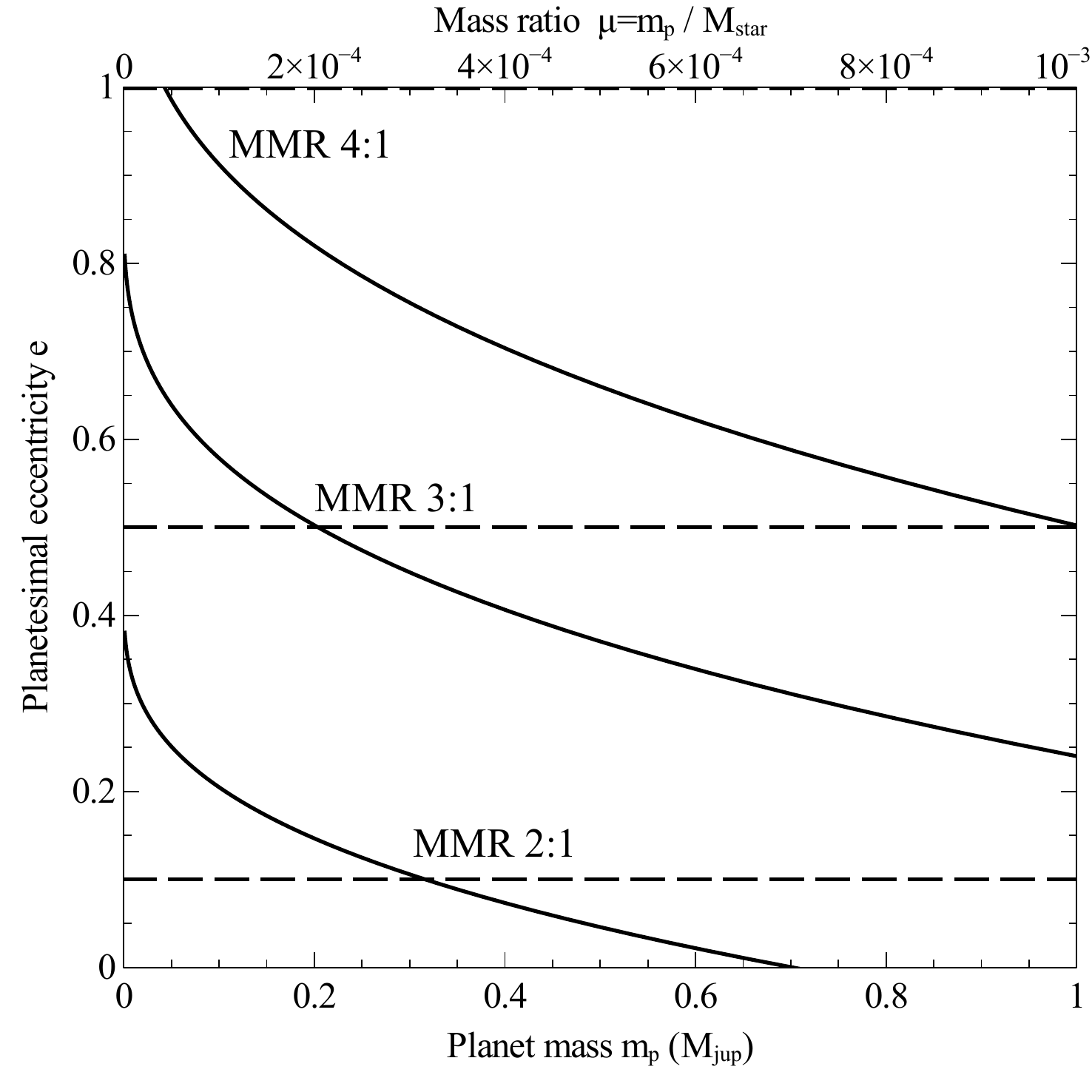}
\includegraphics[width=0.5\textwidth,height=0.5\textwidth]{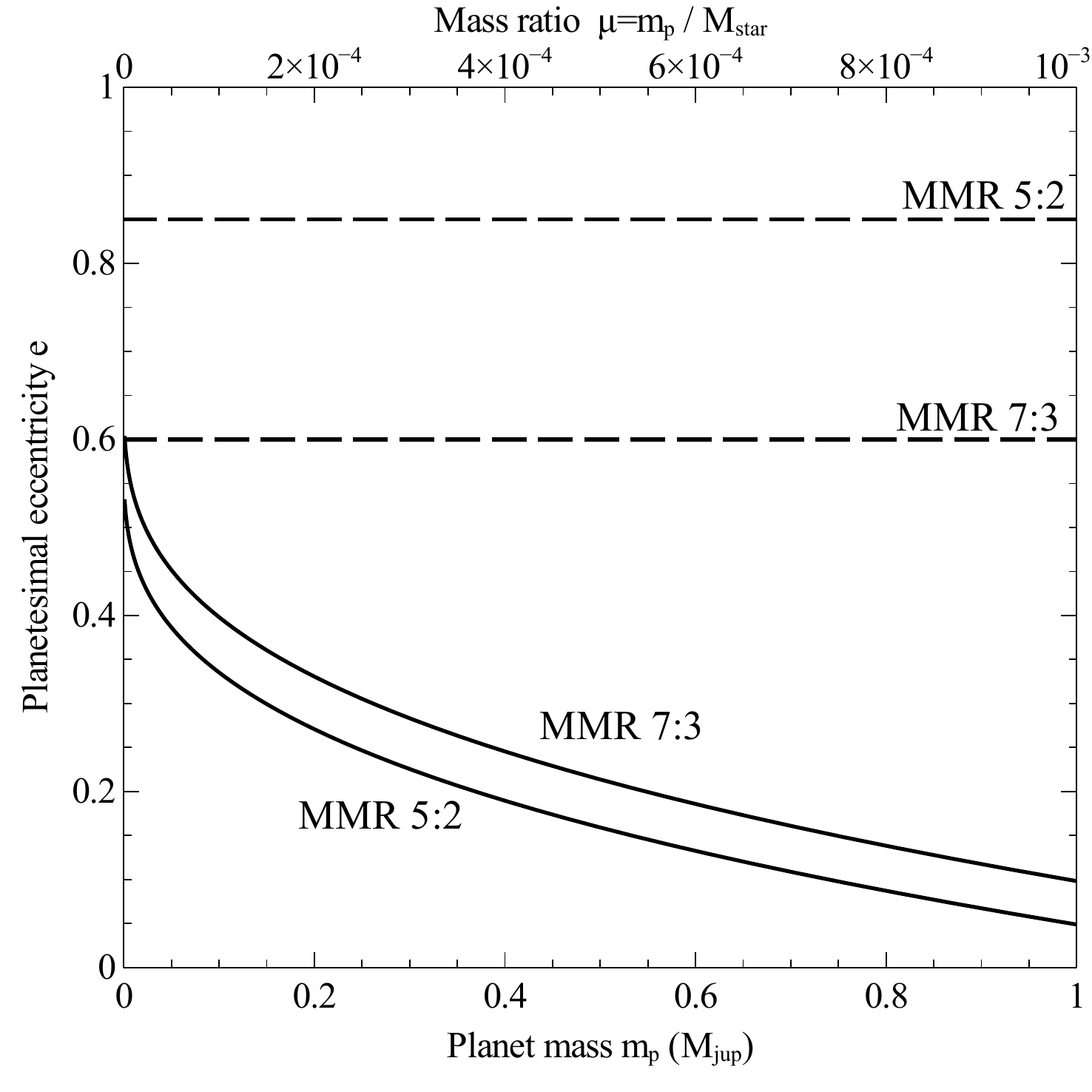}}
\caption[]{The solid lines show the critical eccentricity $\mathrm{e_{crit}}$ above which a planetesimal should be driven in order to cross the chaotic zone of a planet (solid lines), as a function of the mass ratio between the planet and the star, and for the 4:1, 3:1, 5:2, 7:3, and 2:1 MMRs. These results are independent from the semi-major axis of the planet, according to Eq.~(\ref{eq:ability2}), and hold for a planetary orbital eccentricity $\mathrm{e_p}=0.1$. The axis showing the mass of the planet is obtained by assuming $\mathrm{M_{\star}=1\,M_{\odot}}$. The horizontal dashed lines show the maximum eccentricity $\mathrm{e_{max}}$ that a given MMR can induce on a planetesimal that has an initial eccentricity $\lesssim 0.05$.}
\label{fig:e_min}
\end{figure*}

As displayed in Fig.~\ref{fig:e_min}, this quantity can be plotted as a function of $\mu$, or $\mathrm{m_p}$ assuming one solar mass star, and a planetary eccentricity of $\mathrm{e_p}=0.1$. If we compare this minimum eccentricity with the range of eccentricities an MMR can induce on a planetesimal, we notice that the 4:1, 5:2, and 7:3 MMRs are expected to be efficient at scattering planetesimals for a large range of planetary masses. The 3:1 MMR should be effective for a smaller range of planetary masses. 

Surprisingly, the 2:1 MMR can be expected to set planetesimals on cometary-like orbits as well, although its phase-space diagram shows that the increase in eccentricity this MMR can induce on a planetesimal is very small. This is due to the proximity of this resonance with the chaotic zone, as seen in Fig.~\ref{fig:MMR_schema}. This case illustrates how the analysis proposed here complements the reading of the phase-space diagram.

Note that a way to retrieve directly the mass regime at which an MMR is supposed to be effective is to consider that $\mathrm{e_{max}}$ is independent of the planet mass, and can be approximated and fixed for a given $\mathrm{n:p}$ MMR thanks to the corresponding phase-space diagram. One can then derive a minimum planet-star mass ratio $\mathrm{\mu_{min}}$ above which planetesimals trapped in this MMR will be expected to be able to cross the chaotic zone of the planet. Rewriting Eq.~\ref{eq:ability1} in that sense gives :

\begin{equation}\label{eq:mass_min}
\mathrm{\mu \geq \mu_{min}=(3+e_{\mathrm{p}}) \left[\frac{1}{k}\left( 1- \left(\frac{p}{n} \right)^{2/3} \frac{1+e_{max}}{1-e_p} \right) \right]^3} \qquad.
\end{equation}

An upper limit can be set on the mass of the planet by considering the mass for which the chaotic zone of the planet will overlap with the resonance location, that is $\mathrm{a_{chaos,in}<a_{MMR}}$. Using Eqs.~\ref{eq:MMRloc} and ~\ref{eq:chaosin2}, we find that the maximum planet-star mass ratio $\mathrm{\mu_{p,max}}$ allowed for a planet is :

\begin{equation}\label{eq:mass_max}
\mathrm{\mu \leq \mu_{max}=(3+e_{\mathrm{p}}) \left[\frac{1}{k}\left( 1- \left(\frac{p}{n} \right)^{2/3} \frac{1}{1-e_p} \right) \right]^3} \qquad.
\end{equation}

If the planet exceeds this mass, it will directly and immediately scatter planetesimals on cometary orbits. This would result in an intense and short lived comet scattering, rather than the one necessary to explain the observations of exozodis around old stars. Since the 2:1 MMR is the closer resonance to the chaotic zone, it will have the smallest maximum mass of $0.7\,\mjup$ around a Solar-mass star, while the 7:3, 5:2 and 3:1 are expected to lead to direct scattering for planets with masses 1.3, 1.6, and $2.6\,\mjup$, respectively, and with orbital eccentricity $\mathrm{e_p}=0.1$.

Finally, note that we considered a planetesimal starting to be acted upon when having an almost circular orbit, whereas this is not necessarily the case in reality. Indeed, we assumed that the system has suffered from an instability in order for the perturber at work to have been set onto an eccentric orbit. Therefore, the material remaining in the reservoir belt can be expected to exhibit a certain level of excitation. This means that the planetesimals' initial eccentricities in our mechanism are expected to be actually larger than the value of $0.05$ used throughout this paper. This is not expected to inhibit the mechanism that produces active comets, since, as one can see on phase space diagrams, the greater the initial eccentricity of the particles, the greater the eccentricities it will have access to. However, it could also be expected that the timescale required to reach the necessary high eccentricities and suffer a close encounter with the planet is reduced, hence leading to a more rapid production of comets.

%
\section{Numerical study}\label{sec:results}
Studying the behaviour of a swarm of planetesimals by N-body simulations is an excellent way to complement our analytical study. It will allow us to explore statistics and timescales of the mechanism that sets planetesimals on cometary orbits.
In this section where we focus on the 5:2 MMR as a study case, we first describe the method we followed (code, initial conditions, parameters explored), before describing the results of our numerical simulations.

\subsection{Numerical initial conditions}
 
Our initial sets of planetesimals all contain 5,000 massless test particles, for which eccentricity and inclination are randomly and uniformly distributed between 0-0.05 and $\pm 3^{\circ}$, respectively, which correspond to what can be expected from a cold debris disc, as constrained from observations of the Kuiper belt where $\mathrm{e_{max}}\sim 0.1$ and $\mathrm{i_{max}}\sim 3^{\circ}$ \citep{2011AJ....142..131P}.
The value of 0.05 chosen for the maximum initial eccentricity of our particles is arbitrary compared to the 0.1 limit of \citet{2011AJ....142..131P}.
The remaining characteristic angles of the test particles, the longitude of ascending node $\Omega$, the longitude of periastron $\omega$ and the mean anomaly $\mathrm{M}$, are all randomly and uniformly distributed between 0 and $2\pi$, that is, the orbits exhibit no preferential direction. 
Their semi-major axis distribution spans 0.5 AU, among which the particles semi-major axes are randomly and uniformly distributed. These intervals are centred on the exact mean-motion resonance location, as given by Eq.(\ref{eq:MMRloc}). Their location will depend on the MMR and on the semi-major axis of the planet, therefore, the location of our initial ring of planetesimals will change from one simulation to another. As seen in the previous section, analytical studies indicate that the architecture of the system scales with $\mathrm{a_p}$. Therefore, this parameter is expected to have no influence on the feasibility of the process.

In other words, it is only the mass of the planet that is expected to determine the feasibility, while the scattering rates, as well as delays and duration, are expected to depend on both the mass and the semi-major axis of the planet.
Larger semi-major axes, that is, a larger scale of the architecture, translates into longer periods, and thus larger dynamical timescales. On the other hand, dynamical timescales are expected to be larger at smaller $\mathrm{m_p}$, since less massive planets are expected to make planetesimals travel slower across their phase-space. 
We explore planetary masses $\mathrm{m_p}=0.1-0.25-0.5-0.75-1\,\mjup$ and semi-major axes $\mathrm{a_p}=25-50-75-100\,$AU. The influence of the mass of the central star will be discussed in Sect.~\ref{sec:applis} as well as its luminosity.
In all cases, the central star is $\mathrm{M}_{\star}=1\,\mathrm{M}_{\odot}$, and we only consider the case where there is no mutual inclination between the ring and the orbit of the planet, that is, the orbital inclination of the planet is always zero. It is certainly not realistic regarding our assumptions on the origin of the eccentric perturber, because planets are expected to suffer changes in both eccentricity and inclination when being scattered.

Determining the effect of mutual inclination would imply carrying out a 3D analysis for the 5:2 MMR, as was done for the 4:1 MMR by \citet{2000Icar..143..170B} and \citet{2007A&A...466..201B}. This is beyond the scope of this paper, however, note that \citet{2000Icar..143..170B} showed that mutual inclination of $(\lesssim 10^{\circ})$  between the orbit of an eccentric planet and a reservoir in 4:1 MMR does not inhibit the generation of active comets via direct placement\\ \citep{2000Icar..143..170B}, as their results in the coplanar case and the case with a mutual inclination were almost identical.

We integrate the evolution of this system over 1 Gyr, using the symplectic N-body code SWIFT-RMVS \citep{1994Icar..108...18L}. We used a typical time-step of $\sim 1/20\,$ of the smallest orbital period, which ensures a conservation of energy with a typical error of $\sim 10^{-6}$ over the whole integration, since the symplectic nature of the code ensures there is no long-term drift of energy error. Particles were removed when approaching the star at a distance of less than $\sim 1 \mathrm{R}_{\odot}$. We record snapshots of our simulations every 1 Myr and for each snapshot, we examine the particles which have orbits with $\mathrm{q<q_{sub}}$ and calculate how many pericentre passages they have per year. We refer to the total number of pericentre passages per year of all active comets as the cometary events per year.

\subsection{Results}

A large proportion of our simulations gave rise to what was expected, that is a delayed production of active comets, and then a continuous production at an approximately constant rate, until the system reaches 1 Gyr. This continuous production of cometary orbits was expected, because planetesimals may take several precession cycles and cross the chaotic zone of the planet several times before being scattered on to a comet-like orbit. Also, as mentioned in the previous section, the delay gives us an estimate of the precession timescale. Therefore, it was also expected that in the case of large ($\sim 100\,$Myr) delays, the production of active comets would be maintained for at least a few precession cycles and therefore until 1 Gyr. However, in some cases the delay is very short, that is inferior to 50 Myr, and yet the activity is still maintained up to 1 Gyr. Consequently, the number of cycles necessary for all the planetesimals in MMR to encounter the planet is of the order of several tens.

\begin{figure}
\centering
\includegraphics[width=0.55\textwidth]{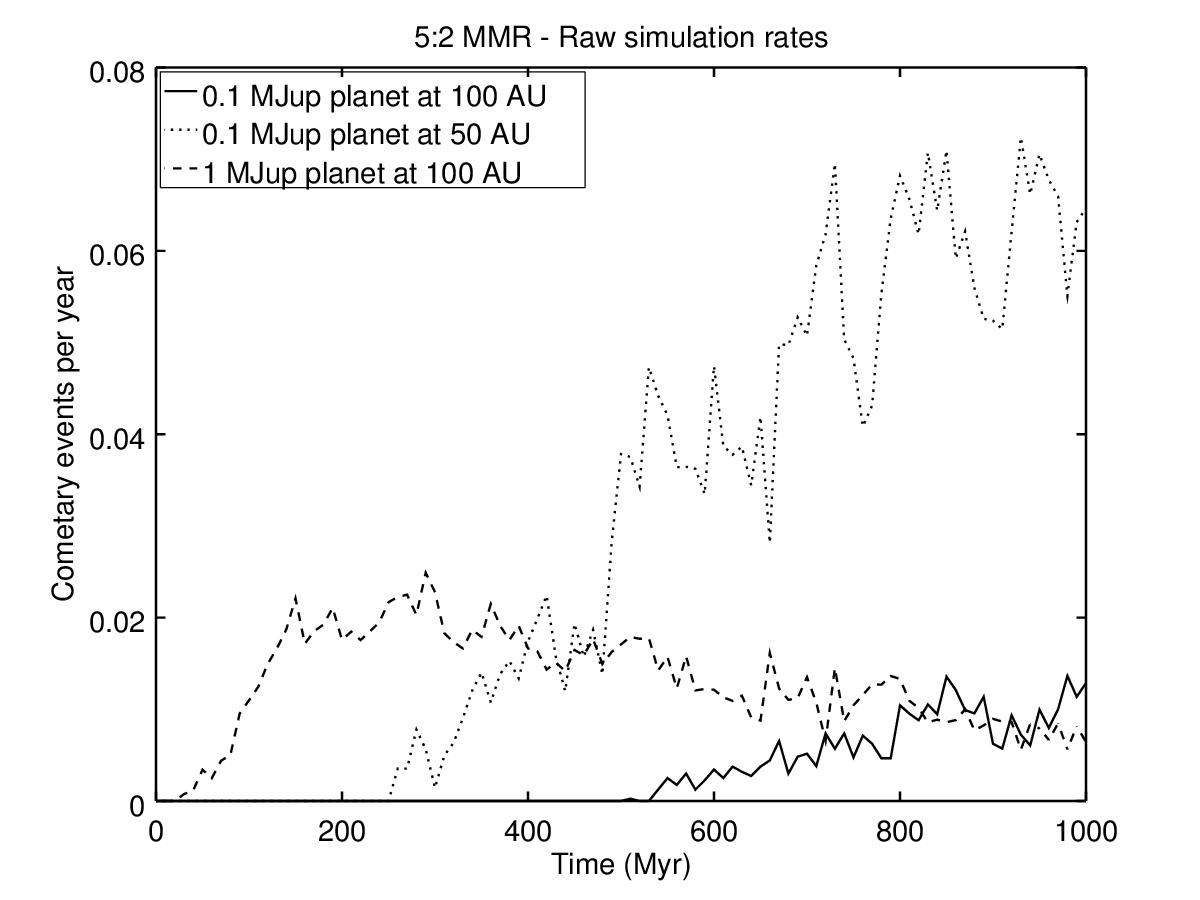}
\caption[]{Evolution of the number of cometary events per year as a function of time, when the cometary event is generated via the 5:2 MMR with a planet of eccentricity $\mathrm{e_p=0.1}$ with planet of mass $0.1\,\mjup$ at 100 AU \emph{(solid line)}, a planet of mass $0.1\,\mjup$ at 50 AU \emph{(dashed line)}, and a planet of mass $1\,\mjup$ at 100 AU \emph{(dotted line)}. The delay increases with increasing semi-major axis and decreasing mass, as expected. The production of active comets can be sustained up to at least 1 Gyr and probably beyond, even when the delay is short. Note that the rates presented here are raw rates from our numerical simulations that shall be translated into more realistic values in Sect.~\ref{sec:applis}, and that the curve is time-averaged over 10 Myr intervals to reduce some of the stochasticity that results from the resolution of the simulations.}
\label{fig:example_rate}
\end{figure}

In Table~\ref{tab:simus}, we summarise the results of the delay in the appearance of active comets. These delays range from several 10 Myr to several 100 Myr. As expected, this timescale increases with decreasing mass of the planet (see Fig.~\ref{fig:example_rate}). It also increases for increasing semi-major axes, however, there are exceptions to this rule, for which very few comets were produced and/or the timescales do not match the expected trend. These exceptions occur for planets with masses $\leq 0.5\,(\mjup)$, and at 75 AU, which appears to contradict the expectation that the semi-major axis has no impact on the feasibility of the process. Although we do not understand yet what is the cause of these irregularities, studying the problem will probably require an extensively detailed analysis of the orbital evolution of the test-particles. This will be the subject of future work and we will focus here on the most common behaviour.

\begin{figure}
\centering
\includegraphics[width=0.55\textwidth]{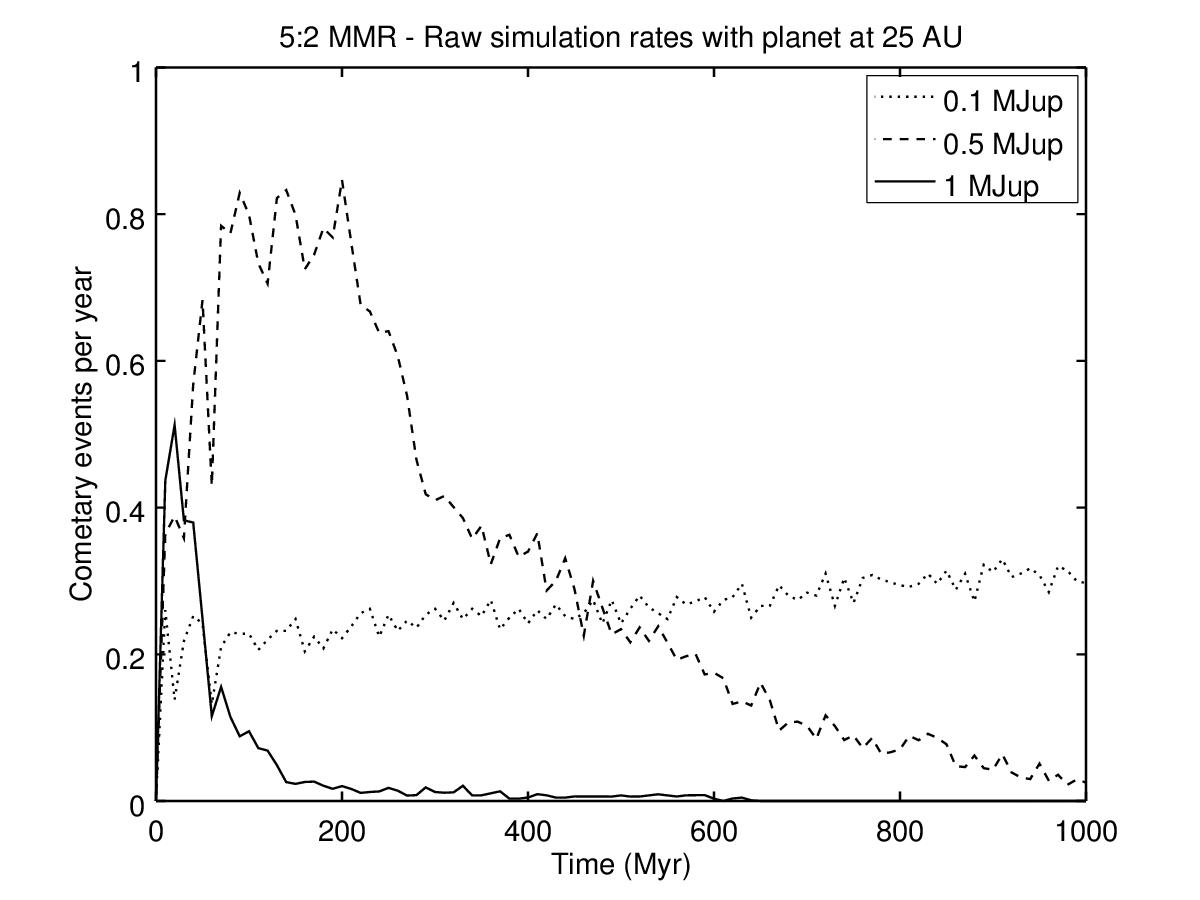}
\caption[]{Evolution of the number of cometary events per year as a function of time, when the cometary event is generated via the 5:2 MMR with a planet of eccentricity $\mathrm{e_p=0.1}$ at 25 AU, and with mass $0.1\,\mjup$ (\emph{dotted line}), $0.5\,\mjup$ (\emph{dashed line}), and $1\,\mjup$ (\emph{solid line}). Here the placement of planetesimals on cometary orbits is produced directly by the MMR without the need for an additional scattering event. Therefore, as expected, planetesimals are set on cometary orbits in only few Myr, and the activity profile is more intense. If the planet has a mass of the order of that of Jupiter, the production of cometary orbits is sustained on timescales shorter than in the case the production is indirectly induced via scattering events. Note that it can nevertheless be maintained on timescales of the order of several 100 Myr. In the case of less massive planets, direct production can be sustained up to at least 1 Gyr and probably beyond.}
\label{fig:direct}
\end{figure}

The cases where a planet is at 25 AU appear to be limiting cases. In these cases the production of active comets is almost immediate with delays of several Myr only. This was expected because at this semi-major axis, as we mentioned in the previous section, the resonance will sustain direct production of cometary orbits, without the need for an additional scattering event. We also expected a profile with a sharp peak of intense production of active comets, which we indeed observe for planets with masses $\mathrm{m_p}\geq 0.5\,(\mjup)$ (see Fig.~\ref{fig:direct}). As expected, the peak rate is at least one or two orders of magnitude larger than when the planet is at larger distances than 25 AU. Note however, that although the activity is not maintained until 1 Gyr, it was maintained for several 100 Myr, which may be sufficient to produce an exozodi in systems younger than this. Interestingly, at the same distance, planets with masses $\mathrm{m_p}\leq 0.5\,(\mjup)$ can cause the direct production of active comets to be sustained for Gyr timescales and with rates comparable or greater than those obtained via the indirect mechanism (see Fig.~\ref{fig:direct}).

In the remaining cases, the number of cometary events per year spans two orders of magnitude with a tendency for this rate to be smaller for planets on very distant orbits than for closer planets. This tendency can be easily explained, as comets that originate at larger distances have larger semi-major axes, and longer periods, which naturally diminishes the number of cometary events per year.
The dynamical lifetime of a comet-like orbit is in general inferior to 1 Myr (in $\sim 75\%$ of orbits), with a tendency for smaller mass planets to give the largest survival timescales.

However, the values numerically found for the rates and survival lifetime must be taken with caution, because they are not necessarily realistic. We used 5,000 test particles in our simulations while real belts contain much more bodies. Besides, the dynamical lifetime of a cometary orbit may differ substantially from the actual lifetime of the comet against evaporation. 

\begin{table*}
 \caption{Properties of the simulations as a function $\mathrm{a_p}$, and $\mathrm{m_p}$:  peak cometary event rate $\mathrm{N_{max}}$, delay before active comets are seen $\mathrm{t_{del}}$ in Myr, duration of the production of active comets $\mathrm{t_{prod}}$ in Myr, or minimum duration if the production is still ongoing at the end of our simulations, and FWHM of the peak in Myr, or minimum FWHM in case the time for the profile to reach again half of the peak value exceeds the duration of our simulations. Note that in some cases, it is questionable whether the maximum rate is reached during the simulation. There were no active comets produced for a planet of 0.1 $\mjup$ at 75 AU.}
\label{tab:simus}

\begin{tabular*}{\textwidth}{@{\excs}c|cccccc}
\toprule
  \multirow{2}{*}{$\mathrm{a_p}$ (AU)}   & & \multicolumn{5}{c}{$\mathrm{m_p}\,(\mjup)$} \\[5pt]
  & & 0.1 & 0.25 & 0.5 & 0.75 & 1 \\[5pt]
\midrule
    & $\mathrm{N_{max}}$ & $\sim 3.3\times 10^{-1}$ & $\sim 6.2\times 10^{-1}$ & $\sim 8.5\times 10^{-1}$ & $\sim 7.9\times 10^{-1}$ &  $\sim 5.1\times 10^{-1}$ \\[2pt]
  25  & $\mathrm{t_{del}}$ & $ \sim$ 5   & $\sim$ 5   & $\sim$ 5   & $\sim$ 5  & $\sim$ 5\\[2pt]
    & $\mathrm{t_{prod}}$&$\gtrsim$ 995 & $\gtrsim$ 995 & $\gtrsim$ 995 & $\gtrsim$ 995 &  $\sim$ 600 \\[2pt]     
    & FWHM & $\gtrsim$ 900 & $\sim$ 900 & $\sim$ 250 & $\sim$ 100 &  $\sim$ 50 \\[2pt]   
\midrule
    & $\mathrm{N_{max}}$&$ \sim 7.2\times 10^{-2}$ & $\sim 8.6\times 10^{-2}$ & $\sim 8.0\times 10^{-2}$ & $\sim 8.9\times 10^{-2}$ &  $\sim 9.6\times 10^{-2}$ \\[2pt]
50  & $\mathrm{t_{del}}$&$ \sim$ 250 & $\sim$ 60  & $\sim$ 40  & $\sim$ 30 & $\sim$ 20    \\[2pt]
    & $\mathrm{t_{prod}}$&$ \gtrsim$ 750 & $\gtrsim$ 940 & $\gtrsim$ 960 & $\gtrsim$ 970 &  $\gtrsim$ 980 \\[2pt]     
    & FWHM & $\gtrsim$ 500 & $\gtrsim$ 700 & $\sim$ 900 & $\sim$ 700 &  $\sim$ 300 \\[2pt]  
\midrule
    &  $\mathrm{N_{max}}$& &$  \sim 3.5\times 10^{-3}$ & $\sim 3.3\times 10^{-2}$ & $\sim 4.1\times 10^{-2}$ &  $\sim 2.6\times 10^{-2}$ \\[2pt]  
75  &   $\mathrm{t_{del}}$& &$  \sim$ 600 & $\sim$ 200 & $\sim$ 60 & $\sim$ 30   \\[2pt]
    &   $\mathrm{t_{prod}}$& &$ \gtrsim$ 400 & $\gtrsim$ 800 & $\gtrsim$ 940 &  $\gtrsim$ 970 \\[2pt]     
    &  FWHM & & $\gtrsim$ 300 & $\gtrsim$ 400 & $\sim$ 800 &  $\sim$ 600 \\[2pt]  
\midrule
    & $\mathrm{N_{max}} $&$\sim 1.4\times 10^{-2}$ & $\sim 2.1\times 10^{-2}$ & $\sim 3.6\times 10^{-2}$ & $\sim 3.7\times 10^{-2}$ &  $\sim 2.5\times 10^{-2}$ \\[2pt]
100 & $\mathrm{t_{del}} $&$\sim$ 500 & $\sim$ 200 & $\sim$ 100 & $\sim$ 70 & $\sim$ 40  \\[2pt]
    & $\mathrm{t_{prod}}$&$\gtrsim$ 500 & $\gtrsim$ 800 & $\gtrsim$ 900 & $\gtrsim$ 930 &  $\gtrsim$ 960 \\[2pt]     
    & FWHM & $\gtrsim$ 200 & $\gtrsim$ 500 & $\gtrsim$ 700 & $\gtrsim$ 750 &  $\sim$ 700 \\[2pt]  
\bottomrule     
\end{tabular*}

\end{table*}

%
\section{Applications}\label{sec:applis}

In this section, we apply our findings to specific cases and show how the raw rates from our simulations can be rescaled and interpreted for different situations. This will allow us to discuss the influence of the mass of the initial reservoir and of the star's luminosity. In particular, we will investigate the case of a low mass disc around a Solar-type star and demonstrate that even a low mass disc is a reservoir that can provide enough active comets to sustain the amount of dust necessary to produce the detected exozodis. The second application demonstrates how high the comet scattering can be when considering a much larger disc around an A-type star, namely Vega.
\subsection{Rescaling raw rates}

Firstly, the lifetime of a comet against evaporation $\mathrm{t_{evap}}$ may differ from the survival dynamical timescale $\mathrm{t_{dyn}}$ of its orbit. If comets evaporate faster than their dynamical timescale, the raw rates will give an overestimation and should therefore be corrected by a factor $\mathrm{t_{evap}/t_{dyn}}$. 
The dynamical timescale can be retrieved from our simulations, while $\mathrm{t_{evap}}$ can be determined using the recent cometary evaporation models of \citet{2016arXiv160403790M}.

Secondly, our simulations contain $\mathrm{N_{init}}=5,000$ test particles distributed over a radial range of 0.5 AU, which is extremely small compared to what would be expected in real debris discs.
For instance, in the Kuiper belt interior to 50 AU, there is an estimate of $5\times 10^9$ comets \citep{1995AJ....110.3073D}. The principal part of the Kuiper belt extends between the 2:3 and 1:2 MMR with Neptune, that is from 39.5 to 48 AU, which gives $\sim 10\,$AU in radial extension.
Therefore, if our simulations were to reproduce realistically a Kuiper-belt analogue, these would have to contain $\mathrm{N_{real}}=2.5 \times 10^8$ test-particles.
It implies that the raw production rates obtained from our simulations have to be rescaled in order to obtain realistic values. As was expected, and as checked numerically by running additional simulations with a planet at 100 AU and with mass $0.25\,\mjup$, and with 1,000, 2500, and 10,000 test-particles, the rate scales linearly with the initial number of particles in the belt.
Consequently, realistic rates are obtained by correcting the raw rates by a factor $\mathrm{\frac{N_{real}}{N_{init}}}$, and which value is $5\times 10^4$ in the case the reservoir is a Kuiper belt analogue.
\subsection{Interpretation and comparison with observations}

Comparing our results with observations of hot exozodiacal dust around other Sun-like stars will rely on the assumption that we are looking at exact analogues to our own Solar System zodiacal cloud, and considering bodies sublimating below a 3 AU distance from their host star. 
This approach is limited since the detected exozodis can be made of dust very close to the star (and therefore hot). Indeed, if we want to test in a direct way whether our mechanism can reproduce the amount and incidence rate of this dust, we must measure how much dust can be brought close enough to the star and 3 AU for a main sequence star is not necessarily close enough.
However, the need for this assumption arises from the fact that it is difficult to derive the dust mass present in the detected exozodis. Indeed, it requires well constrained dust properties which are not available. On the other hand, the near-IR emission of our own Zodiacal cloud is poorly constrained, and strongly relies on models. Therefore, without this assumption, we would not be able to make any comparison, knowing mass feeding rates of our Zodiacal cloud, but very poorly that of exozodis, or knowning the near-IR emission of exozodis, but very poorly that of the Zodiacal cloud.
Under this assumption, we first estimate the amount of dust that is present in these systems. So far, only a few systems around A type stars have been modeled extensively and dust masses have been derived \citep{2006A&A...452..237A,2011A&A...534A...5D,2013A&A...555A.146L}. The amount of hot dust in these systems is found to be about 100 to 1000 times higher than in our Solar system, where the uncertainties stem from both the modeling uncertainties and the poorly constrained amount of hot dust in our own Solar system \citep{1998EP&S...50..493K,2002Icar..158..360H}. 
We use this number to estimate the amount of dust needed to produce the same ($\sim$1\%) near-infrared dust-to-star flux ratio also observed around Sun-like stars \citep{2013A&A...555A.104A,2014A&A...570A.128E}. We assume that the dust emission in all systems is dominated by thermal emission at the dust sublimation temperature. This is justified, because the interferometric constraints require the dust to be either extremely hot or the emission to be dominated by scattered light \citep{2012A&A...546L...9D,2014A&A...570A.128E}. The latter has been largely ruled out by polarimetric non-detections of the excesses \citep{2016arXiv160408286M} and the confirmation that the majority of the excesses are stable over time scales of years \citep{2016arXiv160805731E}. 
At a given (sublimation) temperature the total dust emission scales with the emitting surface and thus with the amount of dust present if we assume similar dust grain sizes and properties. Given the same flux ratio observed for A type and Sun-like stars, the total dust emission (and thus the dust mass) must scale with the star's emission. A Sun-like star is about an order of magnitude fainter in H band than an A-type star. Under the above assumptions the amount of dust around a Sun-like star required to produce the 1\% flux ratio is about an order of magnitude lower than that required around an A-type star. We thus estimate that the amount of hot dust detected in the near-infrared around Sun-like stars is about one to two orders of magnitude higher than that in our Solar system.
Then, under the assumption that the exozodis observed around Solar-type stars are exact analogs to our Zodiacal cloud, we can estimate the amount of dust released by comets from our simulations by using cometary evaporation models and estimates of the mass feeding rate of the Zodiacal cloud. This amount of dust can then be compared to that estimated from observations (10-100 zodis).
\subsection{Solar-type stars with a Kuiper belt analogue}\label{sec:comp_obs}
\begin{figure}
\centering
\includegraphics[width=0.55\textwidth]{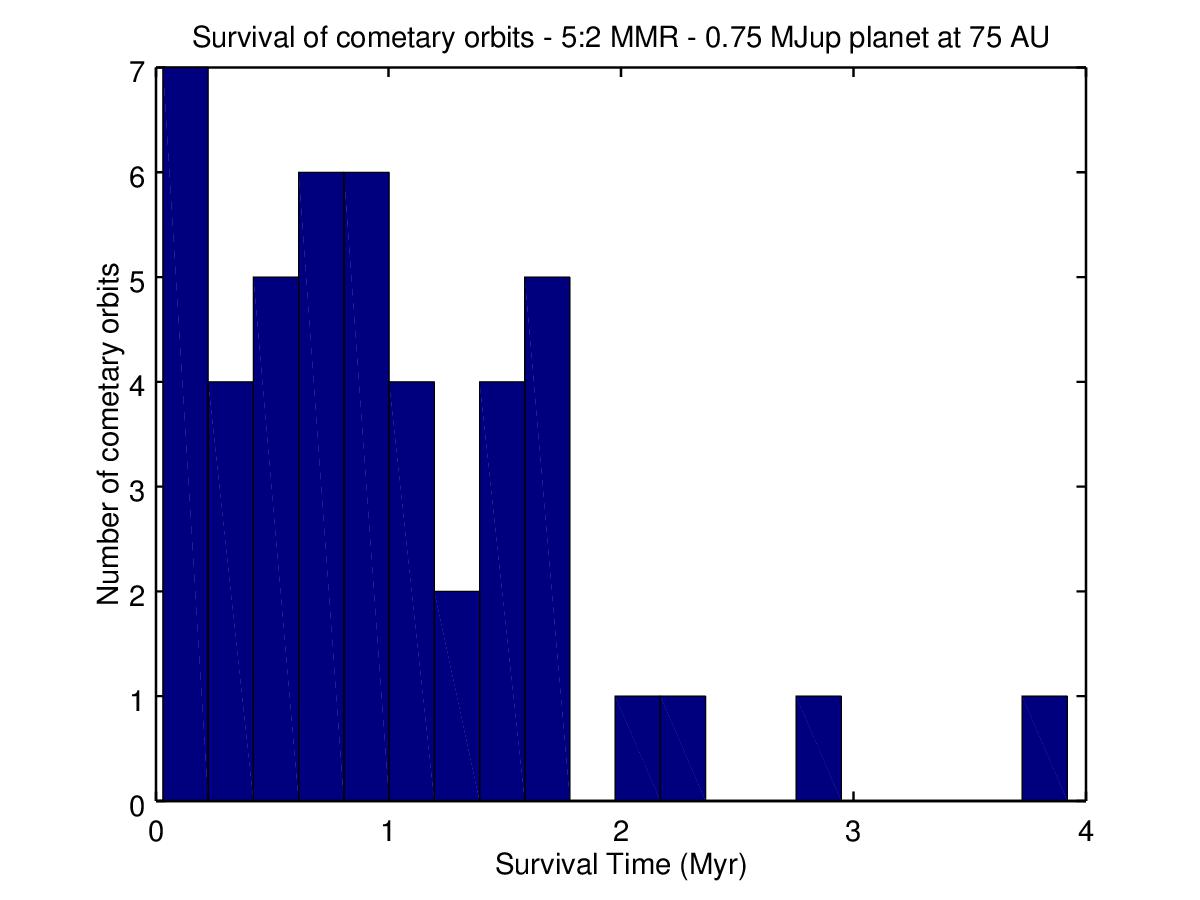}
\caption[]{Histogram of the dynamical survival lifetime of cometary orbits, when the planet involved has eccentricity $\mathrm{e_p=0.1}$, mass $0.75,\mjup$, and is at 75 AU, that is, for a reservoir at the scale of the Solar System Kuiper belt. The majority of the orbits have survival timescales of less than 2 Myr, and of more than several 100,000 yrs, with a mean survival of $\sim 1\,$Myr.}
\label{fig:orbit_life}
\end{figure}

Here we take a look at the case of a Solar type star with a low mass debris disc and consider whether there is enough of a planetesimal reservoir to provide the necessary comets to explain the presence of exozodis around mature stars. To do this, we consider the star to have a disc analogous to the Kuiper belt both in terms of mass and location, but with a planet exterior to it.
\citet{2016arXiv160403790M} found that km-sized bodies survive for 480-4800 orbits around Solar-type stars. From our simulations, $\mathrm{t_{evap}}\sim 0.1-1\,$Myr is the typical amount of time this number of orbits translates into. We retrieved $\mathrm{t_{dyn}}$ by considering our simulation with a planet of mass $1\,\mjup$ at 75 AU, and running an additional 10 Myr integration, taking snapshots every 10,000 yrs. We find that the dynamical lifetime of the orbits is typically 0.1-2 Myr (see Fig.~\ref{fig:orbit_life}), which is consistent with the lifetime of comets against evaporation around Solar-type stars. Consequently, the correction factor is close to, and taken to be unity.
In our simulations, the rates range from $\sim 0.01$ to $\sim 0.05$ cometary events per year. Applying a correction factor $5\times 10^4$ to the raw rates, the realistic rates produced from a Kuiper-belt analogue would be $\sim 500-2500$ cometary events per year.

We wish to compare these rates of $\sim 500-2500$ cometary events per year to the number of active comets that are required in the Solar system to sustain the Zodiacal dust. The rate at which the inner parts of the Solar system are replenished in dust has been estimated to be $10^3\,\mathrm{kg.s^{-1}}$ \citep{1983A&A...118..345L,2010ApJ...713..816N}, that is $\sim 5\times 10^{-15}\,\mathrm{M_{\oplus}.yr^{-1}}$.

\begin{figure}
\centering
\includegraphics[width=0.55\textwidth]{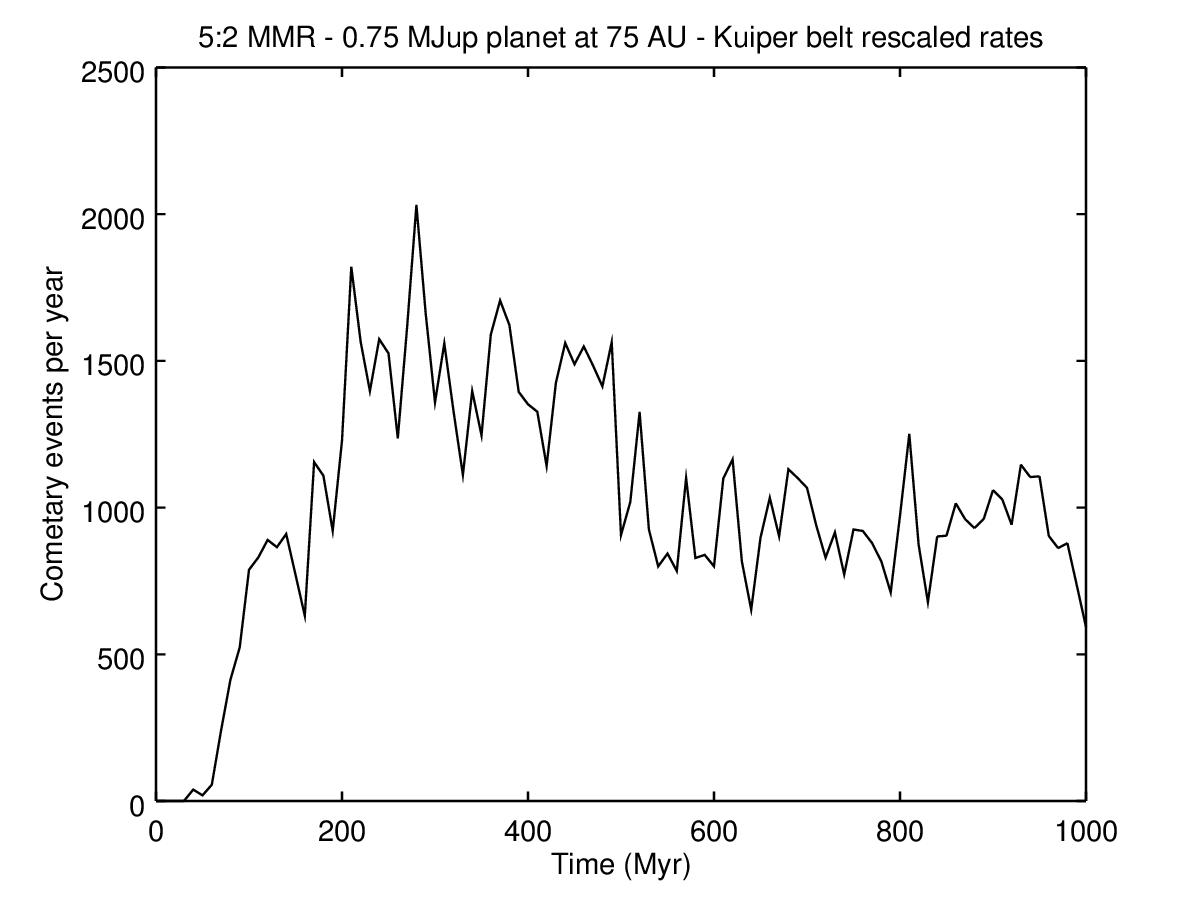}
\caption[]{Evolution of the number of cometary events per year as a function of time, when the cometary event is generated via the 5:2 MMR with a planet with orbital eccentricity $\mathrm{e_p=0.1}$, mass $0.75\,\mjup$ and semi-major axis 75 AU. In this case, the reservoir is at the location of the Kuiper Belt.}
\label{fig:example_rate2}
\end{figure}

Evaporation and sublimation of a typical comet (2.5 km in radius) releases $\sim 3.6\times 10^{-15}\,\mathrm{M_{\oplus}.yr^{-1}}$ on average between 2 and 3 AU \citep{2016arXiv160403790M}, so that the rates of $\sim 500-2500$ cometary events per year from our simulations would translate into mass feeding rates of $\sim 10^{-12}-10^{-11}\,\mathrm{M_{\oplus}.yr^{-1}}$.
If we take for example the case shown in Fig.~\ref{fig:example_rate2}, that is for a planet of mass $0.75\,\mjup$ at 75 AU, the 5:2 MMR would be located at 40.7 AU, which corresponds to the location of the Solar System's Kuiper-belt. In this case, the expected rate is $\sim 1000\,$cometary events per year, that is a few $10^{-12}\,\mathrm{M_{\oplus}.yr^{-1}}$, which is three orders of magnitude greater than the rate at which the Zodiacal cloud is replenished. This means that a planet with an eccentricity of 0.1 exterior to an analogue of the Kuiper-belt in terms of content and scale could produce an exozodi comparable to those that are detected.

Note that according to the results of our simulations, these rates could be also achieved on Gyr timescales thanks to low mass planets $(\lesssim 0.5\,\mjup)$ at distances such that they trigger direct cometary orbits production ($\lesssim 25\,$AU).

These high rates raise the question whether even the massive, hot exozodiacal dust systems detected in the near-infrared \citep{2013A&A...555A.104A,2014A&A...570A.128E} can be produced by our proposed scenario. \citet{2014A&A...570A.128E} found no strong correlation between the presence of hot dust and that of a Kuiper-belt like cold debris disk. Our Kuiper belt would have remained undetected around other stars in available observations due to its low mass \citep[e.g., with Herschel, ][]{2010A&A...520A..32V}. Thus, a scenario that can produce sufficient amounts of hot dust from such a belt could explain this lack of strong correlation.

\subsection{The Vega System}

Extreme feeding rates are required around A-type stars because dust grains have a much shorter lifetime around these stars than around Solar-type stars. A famous example of an A-type star surrounded by an exozodi is Vega. With a luminosity of 40 times that of the sun \citep{2010ApJ...708...71Y}, dust grains have a very short lifetime in its vicinity, of the order of 1 yr due to the short collision times and high radiation pressure \citep{2006A&A...452..237A,2011A&A...534A...5D,2016arXiv160403790M}, whereas grains have lifetimes of $2-6\times 10^{5}$ yrs in the Solar System's Zodiacal cloud due to PR drag \citep{2002ApJ...578.1009F,2003Icar..164..384R}. The quantity of dust deposited by comets must be sufficient to replenish the exozodi on timescales of the order of the grains lifetime. Consequently, the number of comets necessary around Vega can be expected to be more important than it would be to produce a comparable exozodi around a Solar-type star.

\begin{figure}
\centering
\includegraphics[width=0.55\textwidth]{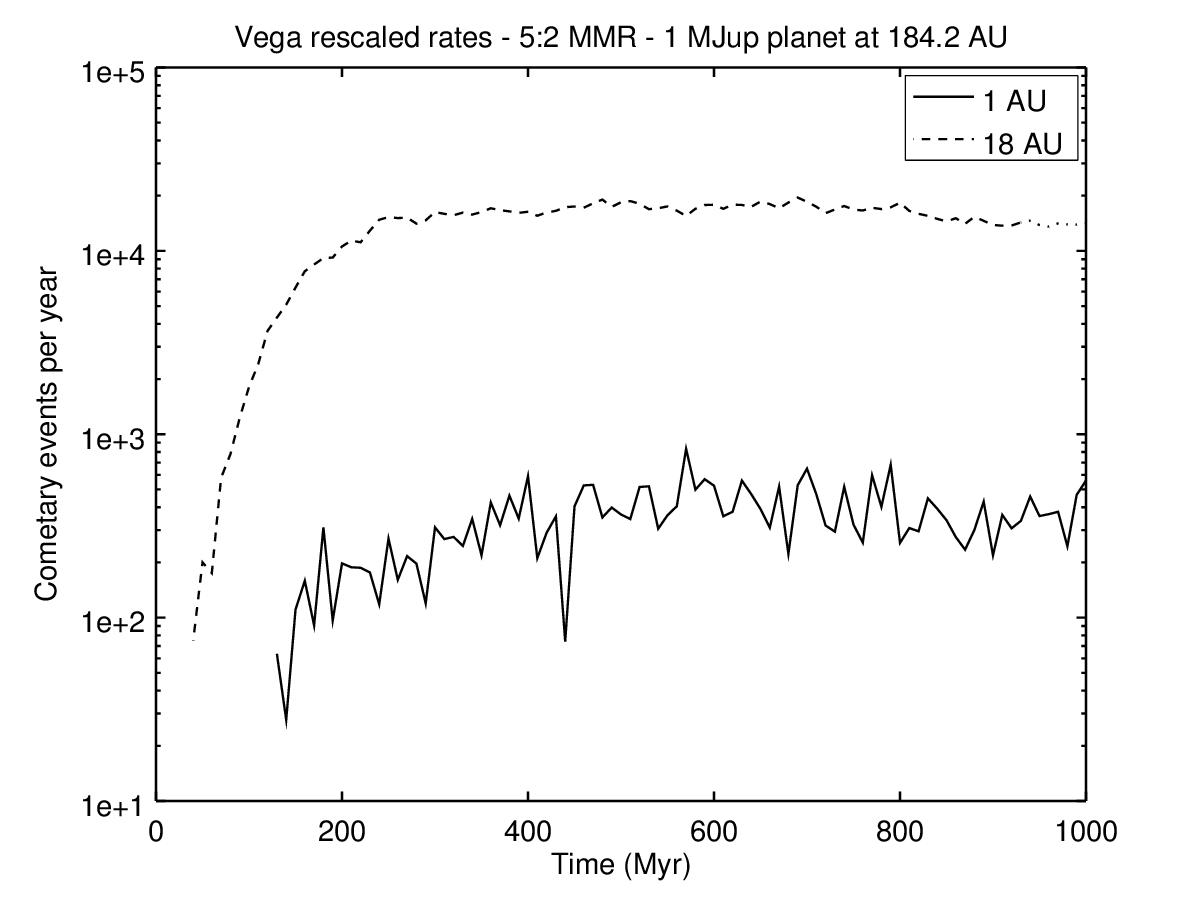}
\caption[]{Evolution of the number of cometary events per year as a function of time, when the cometary event is generated via the 5:2 MMR with a planet of eccentricity $\mathrm{e_p=0.1}$ and semi-major axis 184.2 AU. In this configuration, the 5:2 MMR is located at 100 AU, that is where Vega's cold belt is located. This belt is 500 times more massive than the Kuiper belt so the rate was scaled to what would be expected from such a belt. In addition, the raw rate was also scaled down by a factor 100 because comets are expected to survive on timescales 100 times shorter than their actual lifetime against evaporation around an A-type star. The cometary orbits are defined as orbits with periastron lower than 1 AU on the \emph{left panel}, and 18 AU on the \emph{right panel}.}
\label{fig:Vega_rate}
\end{figure}

Notably, Vega possesses a Kuiper belt analogue which peaks at $\sim 100\,$AU \citep{2005ApJ...628..487S,2010A&A...518L.130S}. Therefore, we ran a simulation specifically dedicated to Vega. According to \citet{2010ApJ...708...71Y}, Vega has a mass of $\mathrm{M_{\star}=2\,M_{\odot}}$. The ring is centred on 100 AU, where it is in 5:2 MMR with an 0.1 eccentric planet, thus located at 184.2 AU. These parameters are summarized in Table \ref{tab:simus3}. We set the mass of the planet to $1\,\mjup$, and otherwise used the exact same conditions as for the rest of our simulations.
Recent cometary evaporation modelling find that the rate required around Vega spans values of 1-2500 cometary events per month, with comets sublimating below 1 AU, that is, where the dust has been observed around Vega \citep{2016arXiv160403790M}. Therefore, we extract from our simulation cometary orbits with periastron smaller than 1 AU. The best-fit mass to the cold belt of Vega is $46.7\mathrm{M_{\oplus}}$ \citep{2010ApJ...708.1728M}, while estimates for the Kuiper-belt give a mass of $0.1\mathrm{M_{\oplus}}$ \citep{2001AJ....122.1051G}, that is 500 times less. Therefore, we use a first correction factor 500 times larger than the one used for the Kuiper belt. Around extremely luminous stars (several tens of $\mathrm{L_{\odot}}$), \citet{2016arXiv160403790M} found that the number of orbits it takes for a comet to evaporate is 4-40. The cometary orbits from our simulations achieve this number of orbits in several 1,000-10,000 years. The comets evaporate on timescales $\mathrm{t_{evap}}$ two orders of magnitude smaller than the typical survival timescale $\mathrm{t_{dyn}}$ of an orbit and a second correction factor of 1/100 is applied to the raw rates. We find that the rate achievable would be $\sim 500$ events per year (see left panel of Fig.~\ref{fig:Vega_rate}). Therefore, the mechanism we present here could sustain several tens of cometary events per month, which is compatible with the rates derived by \citet{2016arXiv160403790M}.
Note that the approach described above supposes that the dust is delivered where the exozodi is observed, which is not necessarily the case. Due to its sensitivity to stellar radiation effects, the dust deposited by comet sublimation is expected to spiral inwards towards the star due to Poynting-Robertson drag effects. On the other hand, the water ice sublimation radius is expected to become larger with increasing stellar luminosity. In the previous sections, cometary orbits had been defined as orbits with periastron smaller than 3 AU, which is relevant around Solar-type stars, however, in the case of an A-type star such as Vega, this sublimation radius is expected to be of the order of $\sim 18$ AU \citep{2016arXiv160403790M}. As can be seen in Fig.~\ref{fig:Vega_rate}, where bodies on orbits with periastrons between 1 and 18 AU are included, the rate of cometary events is one order of magnitude greater. Although PR drag can only bring a limited amount of dust inwards because of competitive collisional processes \citep{2006A&A...452..237A}, the rates extracted from our simulations by considering on site sublimation and disregarding PR drag actually underestimate the real feeding rates.

If this mechanism were at work in the Vega system, the presence of the eccentric planet would have set an imprint on the debris belt, by secularly forcing the eccentricity of the main non-resonant part of the belt to higher values $\mathrm{e_f}$. Applying Eq.(2) of \citet{2014A&A...563A..72F}:

\begin{equation}\label{eq:forced_ecc}
\mathrm{e_f \simeq \frac{5}{4} \frac{a}{a_p} \frac{e_p}{1-e_p^2}} \qquad,
\end{equation}

this would result into the shaping of the belt into a ring with an eccentricity of 0.07, with a center of symmetry that is offset by $\sim 7\,$AU from the star. Such a small eccentricity and offset would not be detectable in the available data. Although it is one of the most famous debris disks and the first one discovered, high angular resolution observations are scarce for this disk. The cool dust was found to be distributed in a radially very broad, face-on disk in mid- to far-infrared observations obtained with Spitzer \citep{2005ApJ...628..487S} and Herschel \citep{2010A&A...518L.130S}. An inner hole is visible in the Herschel/PACS data in the star-subtracted images only \citep{2010A&A...518L.130S,2013ApJ...763..118S}, where the exact stellar location is uncertain. The disk's measured extent seems to be sensitivity limited with its outer edge remaining undetected. In such a situation, measuring an offset of the disk from the central star or detecting a pericenter glow \citep{1999ApJ...527..918W} is not possible. Scattered light or well resolved millimeter interferometric observations are difficult for such a broad disk due to its expected low surface brightness. Clumpy structures in (sub-)millimeter observations of the disk have been reported and attributed to the interaction with a companion \citep{1998Natur.392..788H,2001ApJ...560L.181K,2002ApJ...569L.115W}, but could not be confirmed by more sensitive observations \citep{2011A&A...531L...2P}.Finally, limits from direct imaging observations with Spitzer rule out the presence of planets with mass greater than 2$\,\mjup$ and semi-major axis inferior to 150 AU \citep{2015A&A...574A.120J}.
In summary, the available observations cannot rule out the presence of the planet required in our scenario to explain the production of hot dust in this system. Considering the extent of the debris disc surrounding Vega \citep[several hundreds of AU, ][]{2005ApJ...628..487S}, our scenario implies that the perturbing planet resides within the disc, and at a large separation. This does not seem unrealistic, since planets residing within or even outside a Kuiper belt analogue have been suggested, for instance, as an explanation for a gap in the disc around HD 107146 \citep{2015ApJ...798..124R}, or for a sharp outer edge in the Fomalhaut system \citep{2012ApJ...750L..21B}.

\begin{table}
\caption{Parameters $\mathrm{m_p}$ and $\mathrm{a_p}$, along with the MMR explored and its corresponding location as a function of $\mathrm{a_p}$, for additional simulations. The test-particles are distributed over 0.5 AU centred on the theoretical MMR location computed with Eq.~\ref{eq:MMRloc}, and follow the initial distribution described in Sect.~\ref{sec:results}.}
\label{tab:simus3}

\begin{tabular*}{\columnwidth}{@{\excs}llllll}
\hline\hline\noalign{\smallskip}
Case &  $\mathrm{m_p}\,(\mjup)$ & $\mathrm{a_p}$ (AU) & $\mathrm{e_p}$ & MMR & $\mathrm{a_{MMR}}$ (AU) \\
\hline\noalign{\smallskip}
Vega & 1 & 184.2 & 0.1 & 5:2 & 100\\
2:1 MMR & 0.5 & 100 & 0.1 & 2:1 & 63.0\\
7:3 MMR & 0.5 & 100 & 0.1 & 7:3 & 56.8\\
3:1 MMR & 0.5 & 100 & 0.1 & 3:1 & 48.1\\
$\mathrm{e_p=0.2}$ & 1 & 100 & 0.2 & 5:2 & 54.3\\
\noalign{\smallskip}
\end{tabular*}

\end{table}

\section{Discussion and Conclusion}\label{sec:conclusion}

We found that the 5:2 MMR with a moderately eccentric planet could sustain the observed exozodis on Gyr timescales around Solar and A-type stars. The reservoir belts required are realistic, and could have remained undetected. We took here the example case of the 5:2 MMR, but this process is also expected to occur if the reservoir overlaps with the 7:3, 3:1, and 2:1 MMR. We discuss hereafter the impact of these resonances on the process. We also discuss the impact of the planet eccentricity on the process before drawing our conclusions.

\subsection{Impact of the nature of the MMR}

The rates of cometary events are expected to vary from one MMR to another. Therefore, we use the example case of a planet with an eccentricity of 0.1 at 100 AU, and run additional simulations for the 7:3, 3:1 and 2:1 MMR for comparison with the 5:2. We do not study the 4:1 MMR here, because, as mentioned previously, it would lead to direct placement on cometary orbits without the need for an additional scattering event, no matter the value of the planet semi-major axis. Since it was analytically predicted that the 2:1 MMR would start overlapping with the chaotic zone of the planet for masses greater than $0.7\,\mjup$, we chose to set the mass of the planet to $0.5\,\mjup$. The parameters for these simulations are summarized in Table~\ref{tab:simus3}. 

\begin{figure}
\centering
\includegraphics[width=0.55\textwidth]{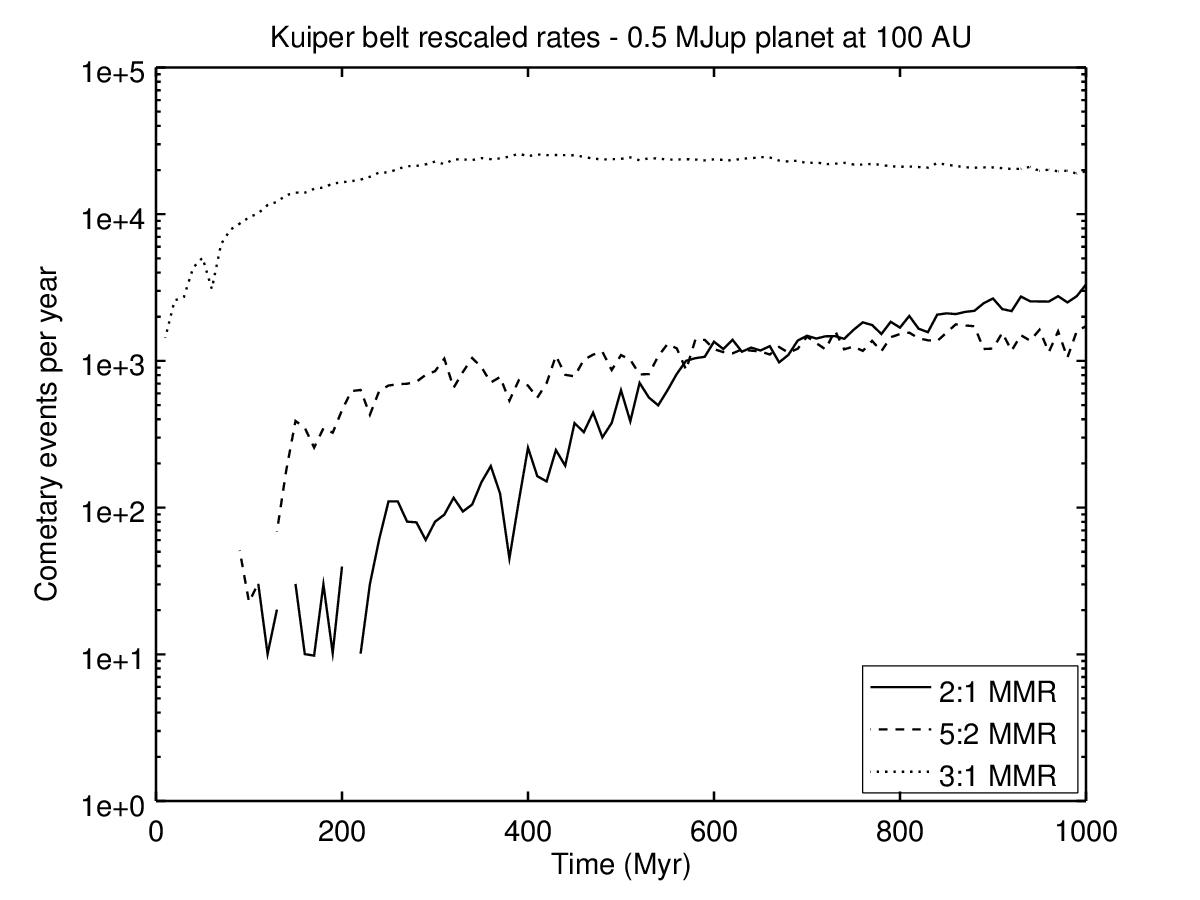}
\caption[]{Evolution of the number of cometary events per year as a function of time, when the production of active comets is generated via the 2:1 \emph{(solid line)}, 5:2 \emph{(dashed line)}, and 3:1 MMRs \emph{(dotted line)}, for a planet at 100 AU, of mass $\mathrm{m_p}=0.5\,\mjup$, and eccentricity $\mathrm{e_p=0.1}$. The rates are given for a reservoir as massive as the Kuiper Belt around a Solar-type star.}
\label{fig:impact_MMR}
\end{figure}

In Fig.~\ref{fig:impact_MMR}, we display the results for the 3:1, 5:2 and 2:1 MMRs, with rates scaled for a reservoir of the mass of the Kuiper Belt as described in Sect.~\ref{sec:applis}. 
The production of active comets generated via the 2:1 MMR shows the same profile as in the case of the 5:2 MMR: it is delayed ($\sim 200\,$Myr), and sustained until 1 Gyr. However, compared with the 5:2 MMR, the rates achieved are several times greater. This means that the exozodis formed via this MMR would be several times more massive than those obtained via the 5:2 MMR.

In the case the planetesimals originate from the 3:1 MMR, the placement on cometary orbits starts almost immediately and the rates are one order of magnitude greater than in the case they originate from the 5:2 or 2:1 MMR. A closer examination of the semi-major axes of the comets formed reveals they are still in MMR while reaching periastrons smaller than 3 AU, which means they have reached eccentricities greater than $\sim 0.95$. This is much larger than the theoretically achievable eccentricity. Indeed, as can be seen on the phase-space diagram, the highest eccentricity expected in this MMR for a planetesimal starting on a low eccentricity orbit (<0.05) is $\sim 0.5-0.6$. Consequently, these results suggest that possible weak planet-particles interactions, e.g., when the planetesimal does not come close enough to the planet to be scattered but suffers nevertheless an impulsion, lead the planetesimals to drift progressively in the phase-space towards higher eccentricities instead of being scattered out of the MMR onto cometary orbits. The 7:3 MMR produced no comets, whereas it was expected to be a channel for cometary production according to analytical predictions in Sect.~\ref{sec:analytical}. We ran an additional simulation for a 0.75 $\mjup$ planet at 75 AU and found the same type of results. A more detailed examination at high resolution of the orbital evolution of test-particles will be needed to understand the behaviour of planetesimals in these two MMRs.

Finally, it is expected that reservoir belts span radial extents of the order of several AU or tens of AU. As shown in Fig. 2 of \citep{1996Icar..120..358B}, the radial width of a resonance behaves as a function of the reduced mass of the planet $\mu$ (where $\mathrm{m_p=\mu (M_{\star}+m_p)}$, which obviously reduces to the mass ratio $\mathrm{m_p/M_{\star}}$ when $\mathrm{m_p << M_{\star}}$). For the cases studied in this paper, this mass ratio is of the order of $\sim 0.001$ or smaller, which corresponds to a radial width of the order of 0.1 AU or smaller. 

Therefore, it could be expected that a realistic reservoir overlaps with several MMRs at the same time. Combining the results found for the case of a 0.5 $\mjup$ planet orbiting at 100 AU, the total rate of cometary events achievable per year would reach values of more than 20,000 when starting from a low mass disc, thus reinforcing the idea that those discs are already sufficiently massive to produce the observed exozodis.

\subsection{Planet eccentricity regime}

We already discussed and explained in Sect.~\ref{sec:analytical} why the process would not work if the planet does not have a sufficient eccentricity. The eccentricity we explored so far, that is $\mathrm{e_p=0.1}$, is considered to be the minimum for the process to take place, however, we did not discuss how this process would be impacted by larger eccentricities.

Firstly, varying the eccentricity of the planet will modify the topology of the phase space, and the planetesimals will be driven to higher eccentricities. In Fig.~\ref{fig:impact_ecc2}, we display the phase space diagram of the 5:2 MMR for planets of eccentricity 0.2 and 0.3. When the planet has an eccentricity of 0.3, it drives directly part of the planetesimals to orbital eccentricities $\sim 1$ and thus, direct comet production is expected in this case, and depending on the mass of the planet, in combination with either comets originating from an additional scattering event or direct scattering.

In addition, when the eccentricity of the planet increases, its periastron diminishes. For a given eccentricity
$\mathrm{e_{p,max}}$, the inner chaotic zone of the planet will start to overlap with the reservoir belt itself, that is for $\mathrm{a_{chaos,in}}\geq \mathrm{a_{MMR}}$. Thus, using Eq.~\ref{eq:MMRloc} and Eq.~\ref{eq:chaosin2}, $\mathrm{e_{p,max}}$ satisfies the equation:

\begin{equation}\label{eq:ecc_cond}
(1-e_{\mathrm{p,max}}) \left[ 1-k\left( \frac{m_{\mathrm{p}}}{(3+e_{\mathrm{p,max}})M_{\star}} \right)^{\frac{1}{3}}\right] - \mathrm{\left(\frac{p}{n} \right)^{2/3}} = 0 \qquad.
\end{equation}

If the planet has an orbital eccentricity that exceeds $\mathrm{e_{p,max}}$, it will directly scatter bodies from the reservoir. It is expected to lead to a short and intense period  of cometary activity.
This condition translates into a maximum eccentricity allowed for the process to take place. For the MMR studied here, $\mathrm{e_{p,max}}$ varies between $\sim 0.2$ and $\sim 0.4$ for planets of mass 0.1-1$\mjup$. Since lower mass planets have smaller chaotic zones, this maximum eccentricity will be larger for lower mass planets. Note that this condition does not depend on the scale of the system. 

\begin{figure}
\centering
\includegraphics[width=0.55\textwidth]{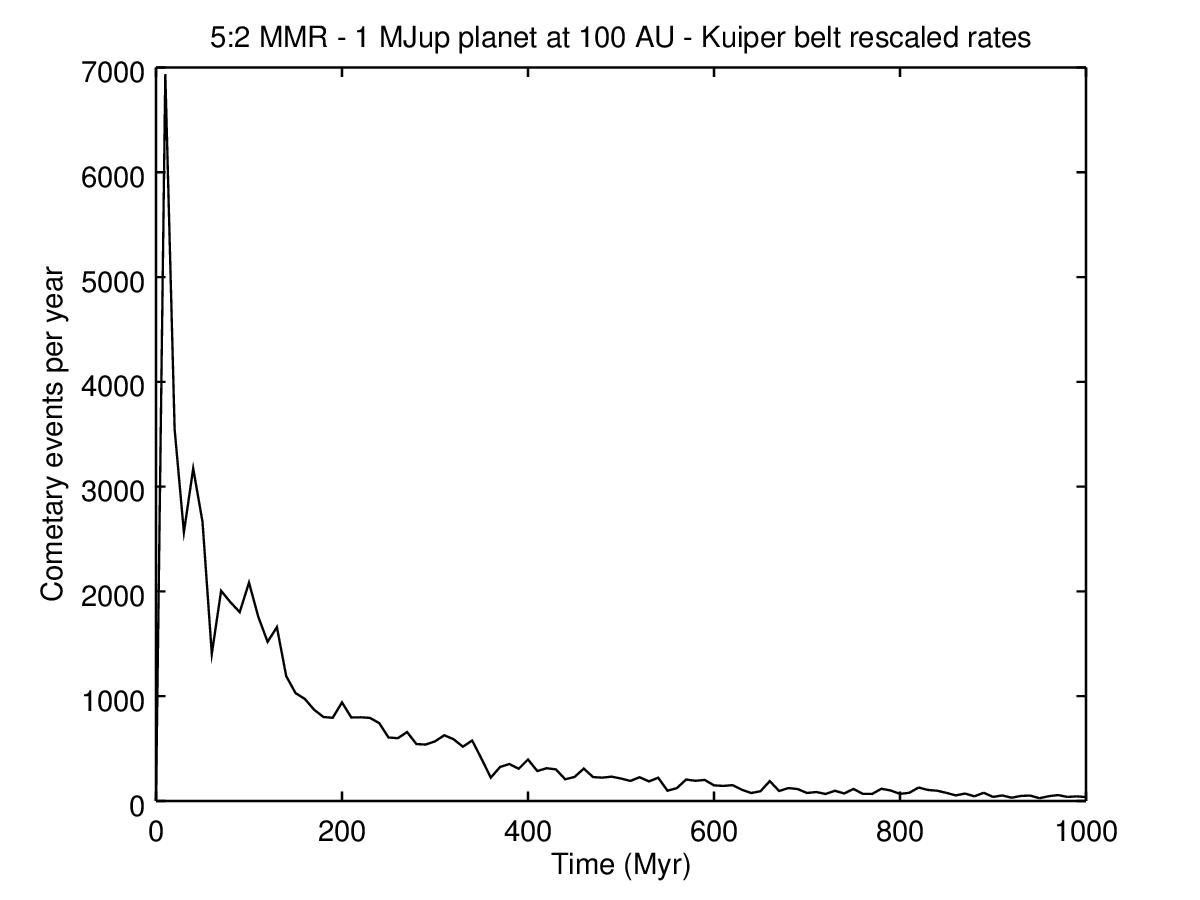}
\caption[]{Evolution of the number of cometary events per year as a function of time, when the production of active comets is generated via the 5:2 MMR for a planet at 100 AU, of mass $\mathrm{m_p}=1\,\mjup$, and eccentricity $\mathrm{e_p=0.2}$. The rates are given for a reservoir as massive as the Kuiper Belt around a Solar-type star.}
\label{fig:impact_ecc1}
\end{figure}

\begin{figure*}
\centering
\makebox[\textwidth]{
\includegraphics[width=0.5\textwidth]{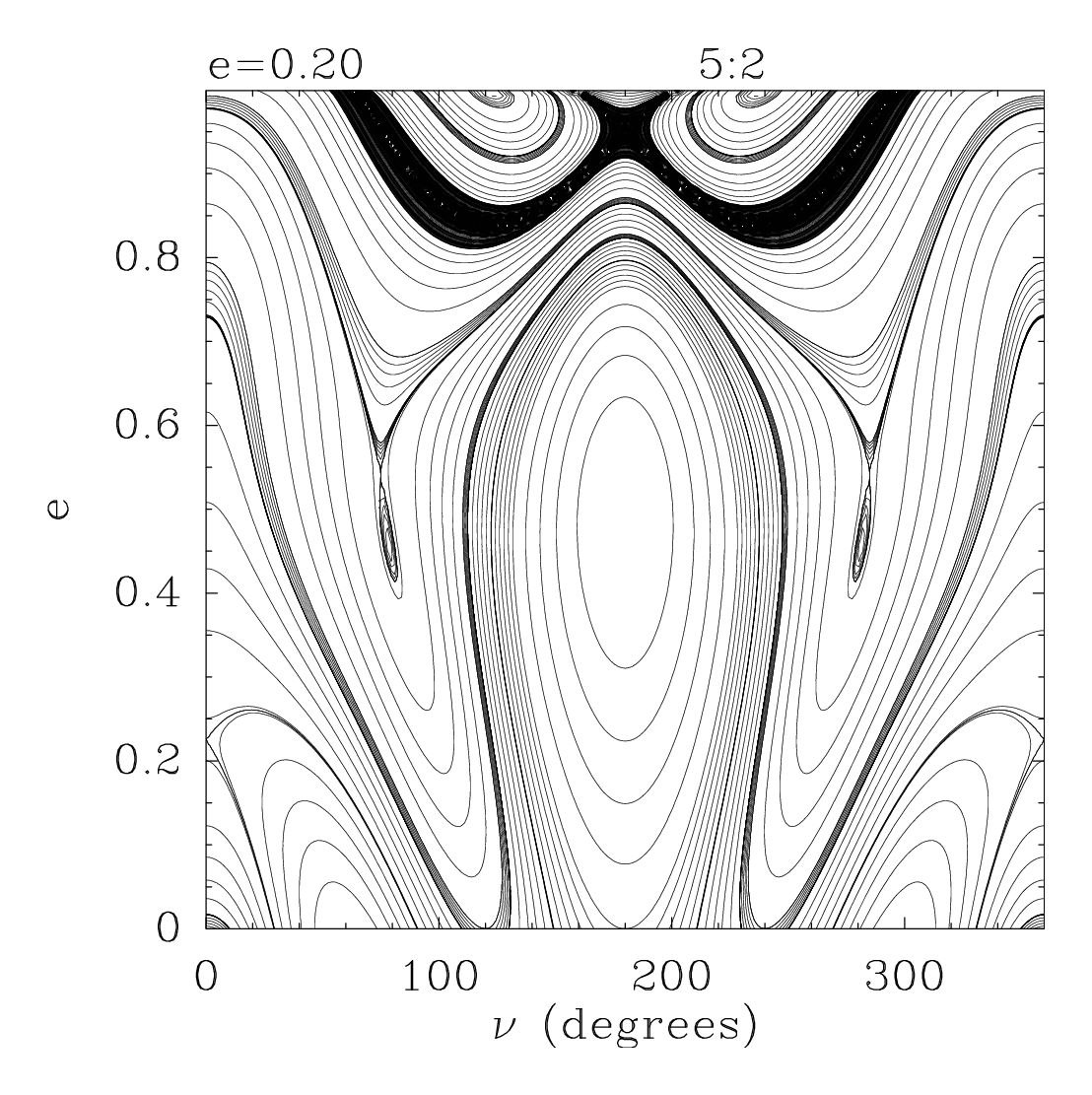}
\includegraphics[width=0.5\textwidth]{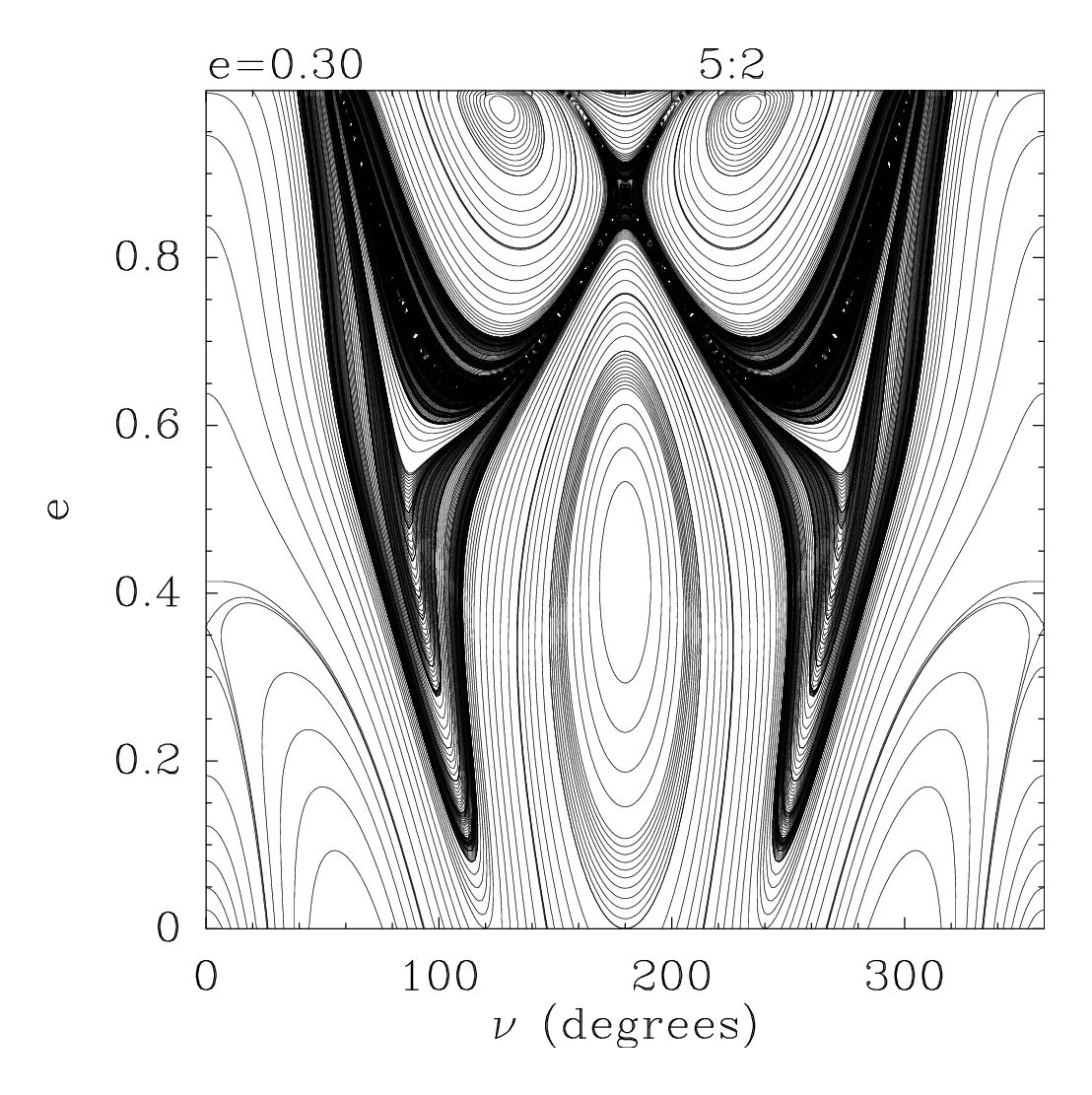}
}
\caption[]{Theoretical phase space diagrams (see caption to Fig.~\protect\ref{fig:phase_space52} for more details) of the 5:2 MMR with a planet of eccentricity $\mathrm{e_p=0.2}$ \emph{(left)} and $\mathrm{e_p=0.3}$ \emph{(right)}.}
\label{fig:impact_ecc2}
\end{figure*}

Using the example where the planet has a mass 1 $\mjup$ and semi-major axis 100 AU, we investigate the effect of the planet eccentricity on the placement of planetesimals on cometary orbits induced via the 5:2 MMR. 
Using Eq.~\ref{eq:ecc_cond}, the eccentricity for which the 5:2 MMR should start inducing direct comet production is $\sim 0.2$. Therefore, we ran the corresponding simulation, and display in Fig.~\ref{fig:impact_ecc1} the time profile of the cometary production. As expected, it is immediate and intense, with rates of several thousands of cometary events per year. It then decays rapidly towards a much smaller rate, maintained until 1 Gyr. This is in accordance with what is expected if a planet directly scatters planetesimals.

\subsection{Conclusions} 

For many reasons, the dynamical mechanism presented here, that is, resonant interactions between planetesimals and an outer eccentric planet followed by scattering on to cometary orbits, is a viable route to produce bright analogues to the Zodiacal Cloud, even in old systems.
Firstly, it can induce production of active comets with potentially significant delays (several 100 Myr). More importantly, it can sustain this comet scattering on Gyr timescales, and at rates compatible with observations when starting from realistic reservoir belts.

We find that when applied around Solar-type stars, our proposed mechanism can sustain a comet production compatible with the formation of the exozodis which are detected, even with low mass reservoirs comparable to the Kuiper Belt. The rate of cometary events that have to be induced around more luminous A-type stars can be sustained as well, and applying our findings to the system of Vega, we suggest that the presence of a planet $\sim 1\,\mjup$ at large distance ($\sim 200\,$AU) and with eccentricity of 0.1 interacting with the system's cold debris disc via MMR could explain the levels of exozodiacal dust detected around the host star.

The mass of the perturbing planet drives the feasibility of the process, while all important quantities at play scale with the semi-major axis of the planet. In other words, if the process occurs for a given planetary mass, it will occur at any scale. Note that at small scales, however, the MMR can induce placement on cometary orbits without the need for an additional scattering event. This is expected to occur with the 5:2 MMR if the planet has a semi-major axis inferior to $\sim 35\,$AU, and inferior to $\sim 15\,$AU for the 7:3 and 3:1 MMRs.
The process is also expected to occur for a large range of planetary masses (typically $0.1-1\,\mjup$) and semi-major axes $(\mathrm{a_p \gtrsim 10-25\,AU})$. The eccentricity regime, $\mathrm{e_p\simeq 0.1-0.4}$, is more restricted.

Our results show that the mechanism is efficient through the 5:2 and 2:1 MMRs. Although predicted to be able to sustain the mechanism, the 7:3 MMR was found to be inefficient, which may be due to its higher order. The 3:1 seems to be particularly efficient, although in this case, comets seem to have formed directly while in MMR thanks to a drift of their eccentricities towards large values. These cases shall be explored in more details in future work, along with the few unexpected irregularities noted in the behaviour of the process with the semi-major axis of the planet. A more extended parametric exploration as well as closer examination of the orbital evolution of test-particles at high temporal resolution should help us explain these cases. 

Inner MMR with eccentric planets is a starting point from which planetesimals can be set on cometary orbits in various ways depending on the planetary mass, eccentricity and semi-major axis regime, as well as on the MMR itself. Planets on short semi-major axes tend to promote direct placement on cometary orbits, for which, interestingly, cases were found for low mass planets ($\lesssim 0.5\,\mjup$) where the production of cometary orbits could be sustained up to 1 Gyr. 
On the other hand, planets at larger separations tend to place planetesimals on cometary orbits indirectly, with a resonant evolution followed by a scattering event. Large planetary eccentricities tend to trigger placement on cometary orbits via direct scattering. Finally, MMR may promote or deny access to cometary orbits via phase space drifts effects.

Some constraints can be drawn on the range of parameters for which each process will take place and how they sometimes superimpose each other. However, they remain crude and several aspects of these processes are for now poorly known and characterised. These aspects shall be the object of future work (Faramaz et al., in prep), performing  extended parametric explorations. 
In particular, it would be useful to be able to predict observational signatures of each of these processes. For instance, with direct placement on cometary orbits while in MMR, planetesimals reach their maximum eccentricity for a specific orbit orientation, so that cometary orbits would have preferential orientations. With the indirect mechanism, additional secular evolution of the planetesimal with the eccentric planet once scattered out of MMR can lead to apsidal alignment with the planet and thus again a preferential orientation of the cometary orbits \citep{2015A&A...573A..87F}. Such preferential alignment of cometary orbits has already been suggested around Beta Pictoris, for which exocomets have been suggested as an explanation for time-variable absorption features in the spectrum of the star, and the lack of variation of the radial velocity has been interpreted as a tendency for cometary orbits to be apsidally aligned \citep{1990A&A...236..202B}. 

Our results suggest that with the mechanism we focused on here, exozodis we detect may originate from as yet undetected cold Kuiper belt analogues, interacting with distant exterior Jupiter-like or smaller planets. This suggests that such systems are likely ideal targets for direct imaging observations. We also note that, whilst we refer in this paper to Kuiper belt analogues, the cometary reservoirs do not necessarily have to be at the outer edge of a planetary system and much of what we discuss here could potentially apply to an asteroid belt analogue. Although we did not explore semi-major axes small enough for the reservoir to be considered an asteroid belt analogue, results for the smallest spatial scale investigated in this study suggest this scenario is worth exploring in more detail, that is, the exozodis that have been detected might instead originate from an MMR with eccentric planets exterior to asteroid belt analogues. 


\section*{Acknowledgements}

The authors wish to thank the anonymous referee for helping us clarify this paper.
VF and MB acknowledge support from FONDECYT Postdoctoral Fellowships, projects nos 3150106 and 3140479.
VF, MB, and JC acknowledge support from the Millenium Nucleus RC130007 (Chilean Ministry of Economy) and Millennium Nucleus "Protoplanetary Disks". MB acknowledges support from the Deutsche Forschungsgemeinschaft (DFG) through project Kr 2164/15-1. JC acknowledges support from CONICYT-Chile through FONDECYT (1141175) and Basal (PFB0609) grants.
CS acknowledges funding by the VRI Summer Research programme for undergraduates of PUC.
The simulations were performed on the Geryon computers at the Center for Astro-Engineering UC (BASAL PFB-06, QUIMAL 130008, Fondequip AIC-57).




\bibliographystyle{mnras}
\bibliography{exozodis} 

\begin{thebibliography}{}
\makeatletter
\relax
\def\mn@urlcharsother{\let\do\@makeother \do\$\do\&\do\#\do\^\do\_\do\%\do\~}
\def\mn@doi{\begingroup\mn@urlcharsother \@ifnextchar [ {\mn@doi@}
  {\mn@doi@[]}}
\def\mn@doi@[#1]#2{\def\@tempa{#1}\ifx\@tempa\@empty \href
  {http://dx.doi.org/#2} {doi:#2}\else \href {http://dx.doi.org/#2} {#1}\fi
  \endgroup}
\def\mn@eprint#1#2{\mn@eprint@#1:#2::\@nil}
\def\mn@eprint@arXiv#1{\href {http://arxiv.org/abs/#1} {{\tt arXiv:#1}}}
\def\mn@eprint@dblp#1{\href {http://dblp.uni-trier.de/rec/bibtex/#1.xml}
  {dblp:#1}}
\def\mn@eprint@#1:#2:#3:#4\@nil{\def\@tempa {#1}\def\@tempb {#2}\def\@tempc
  {#3}\ifx \@tempc \@empty \let \@tempc \@tempb \let \@tempb \@tempa \fi \ifx
  \@tempb \@empty \def\@tempb {arXiv}\fi \@ifundefined
  {mn@eprint@\@tempb}{\@tempb:\@tempc}{\expandafter \expandafter \csname
  mn@eprint@\@tempb\endcsname \expandafter{\@tempc}}}

\bibitem[\protect\citeauthoryear{{Absil} et~al.,}{{Absil}
  et~al.}{2006}]{2006A&A...452..237A}
{Absil} O.,  et~al., 2006, \mn@doi [\aap] {10.1051/0004-6361:20054522}, \href
  {http://adsabs.harvard.edu/abs/2006A%26A...452..237A} {452, 237}

\bibitem[\protect\citeauthoryear{{Absil} et~al.,}{{Absil}
  et~al.}{2013}]{2013A&A...555A.104A}
{Absil} O.,  et~al., 2013, \mn@doi [\aap] {10.1051/0004-6361/201321673}, \href
  {http://adsabs.harvard.edu/abs/2013A%26A...555A.104A} {555, A104}

\bibitem[\protect\citeauthoryear{{Beust} \& {Morbidelli}}{{Beust} \&
  {Morbidelli}}{1996}]{1996Icar..120..358B}
{Beust} H.,  {Morbidelli} A.,  1996, \mn@doi [\icarus]
  {10.1006/icar.1996.0056}, \href
  {http://adsabs.harvard.edu/abs/1996Icar..120..358B} {120, 358}

\bibitem[\protect\citeauthoryear{{Beust} \& {Morbidelli}}{{Beust} \&
  {Morbidelli}}{2000}]{2000Icar..143..170B}
{Beust} H.,  {Morbidelli} A.,  2000, \mn@doi [\icarus]
  {10.1006/icar.1999.6238}, \href
  {http://adsabs.harvard.edu/abs/2000Icar..143..170B} {143, 170}

\bibitem[\protect\citeauthoryear{{Beust} \& {Valiron}}{{Beust} \&
  {Valiron}}{2007}]{2007A&A...466..201B}
{Beust} H.,  {Valiron} P.,  2007, \mn@doi [\aap] {10.1051/0004-6361:20053425},
  \href {http://adsabs.harvard.edu/abs/2007A%26A...466..201B} {466, 201}

\bibitem[\protect\citeauthoryear{{Beust}, {Vidal-Madjar}, {Ferlet}  \&
  {Lagrange-Henri}}{{Beust} et~al.}{1990}]{1990A&A...236..202B}
{Beust} H.,  {Vidal-Madjar} A.,  {Ferlet} R.,   {Lagrange-Henri} A.~M.,  1990,
  \aap, \href {http://adsabs.harvard.edu/abs/1990A%26A...236..202B} {236, 202}

\bibitem[\protect\citeauthoryear{{Boehnhardt}}{{Boehnhardt}}{2004}]{2004come.book..301B}
{Boehnhardt} H.,  2004, {Split comets}.
pp 301--316

\bibitem[\protect\citeauthoryear{{Boley}, {Payne}, {Corder}, {Dent}, {Ford}  \&
  {Shabram}}{{Boley} et~al.}{2012}]{2012ApJ...750L..21B}
{Boley} A.~C.,  {Payne} M.~J.,  {Corder} S.,  {Dent} W.~R.~F.,  {Ford} E.~B.,
  {Shabram} M.,  2012, \mn@doi [\apjl] {10.1088/2041-8205/750/1/L21}, \href
  {http://adsabs.harvard.edu/abs/2012ApJ...750L..21B} {750, L21}

\bibitem[\protect\citeauthoryear{{Bonsor}, {Augereau}  \&
  {Th{\'e}bault}}{{Bonsor} et~al.}{2012}]{2012A&A...548A.104B}
{Bonsor} A.,  {Augereau} J.-C.,   {Th{\'e}bault} P.,  2012, \mn@doi [\aap]
  {10.1051/0004-6361/201220005}, \href
  {http://adsabs.harvard.edu/abs/2012A%26A...548A.104B} {548, A104}

\bibitem[\protect\citeauthoryear{{Bonsor}, {Raymond}  \& {Augereau}}{{Bonsor}
  et~al.}{2013}]{2013MNRAS.433.2938B}
{Bonsor} A.,  {Raymond} S.~N.,   {Augereau} J.-C.,  2013, \mn@doi [\mnras]
  {10.1093/mnras/stt933}, \href
  {http://adsabs.harvard.edu/abs/2013MNRAS.433.2938B} {433, 2938}

\bibitem[\protect\citeauthoryear{{Bonsor}, {Raymond}, {Augereau}  \&
  {Ormel}}{{Bonsor} et~al.}{2014}]{2014MNRAS.441.2380B}
{Bonsor} A.,  {Raymond} S.~N.,  {Augereau} J.-C.,   {Ormel} C.~W.,  2014,
  \mn@doi [\mnras] {10.1093/mnras/stu721}, \href
  {http://adsabs.harvard.edu/abs/2014MNRAS.441.2380B} {441, 2380}

\bibitem[\protect\citeauthoryear{{Booth}, {Wyatt}, {Morbidelli},
  {Moro-Mart{\'{\i}}n}  \& {Levison}}{{Booth}
  et~al.}{2009}]{2009MNRAS.399..385B}
{Booth} M.,  {Wyatt} M.~C.,  {Morbidelli} A.,  {Moro-Mart{\'{\i}}n} A.,
  {Levison} H.~F.,  2009, \mn@doi [\mnras] {10.1111/j.1365-2966.2009.15286.x},
  \href {http://adsabs.harvard.edu/abs/2009MNRAS.399..385B} {399, 385}

\bibitem[\protect\citeauthoryear{{Carpenter} et~al.,}{{Carpenter}
  et~al.}{2009}]{2009ApJS..181..197C}
{Carpenter} J.~M.,  et~al., 2009, \mn@doi [\apjs]
  {10.1088/0067-0049/181/1/197}, \href
  {http://adsabs.harvard.edu/abs/2009ApJS..181..197C} {181, 197}

\bibitem[\protect\citeauthoryear{{Chatterjee}, {Ford}, {Matsumura}  \&
  {Rasio}}{{Chatterjee} et~al.}{2008}]{2008ApJ...686..580C}
{Chatterjee} S.,  {Ford} E.~B.,  {Matsumura} S.,   {Rasio} F.~A.,  2008,
  \mn@doi [\apj] {10.1086/590227}, \href
  {http://adsabs.harvard.edu/abs/2008ApJ...686..580C} {686, 580}

\bibitem[\protect\citeauthoryear{{Chen} \& {Jewitt}}{{Chen} \&
  {Jewitt}}{1994}]{1994Icar..108..265C}
{Chen} J.,  {Jewitt} D.,  1994, \mn@doi [\icarus] {10.1006/icar.1994.1061},
  \href {http://adsabs.harvard.edu/abs/1994Icar..108..265C} {108, 265}

\bibitem[\protect\citeauthoryear{{Crida}, {Masset}  \& {Morbidelli}}{{Crida}
  et~al.}{2009}]{2009ApJ...705L.148C}
{Crida} A.,  {Masset} F.,   {Morbidelli} A.,  2009, \mn@doi [\apjl]
  {10.1088/0004-637X/705/2/L148}, \href
  {http://adsabs.harvard.edu/abs/2009ApJ...705L.148C} {705, L148}

\bibitem[\protect\citeauthoryear{{Defr{\`e}re} et~al.,}{{Defr{\`e}re}
  et~al.}{2011}]{2011A&A...534A...5D}
{Defr{\`e}re} D.,  et~al., 2011, \mn@doi [\aap] {10.1051/0004-6361/201117017},
  \href {http://adsabs.harvard.edu/abs/2011A%26A...534A...5D} {534, A5}

\bibitem[\protect\citeauthoryear{{Defr{\`e}re} et~al.,}{{Defr{\`e}re}
  et~al.}{2012}]{2012A&A...546L...9D}
{Defr{\`e}re} D.,  et~al., 2012, \mn@doi [\aap] {10.1051/0004-6361/201220287},
  \href {http://adsabs.harvard.edu/abs/2012A%26A...546L...9D} {546, L9}

\bibitem[\protect\citeauthoryear{{Duncan}, {Levison}  \& {Budd}}{{Duncan}
  et~al.}{1995}]{1995AJ....110.3073D}
{Duncan} M.~J.,  {Levison} H.~F.,   {Budd} S.~M.,  1995, \mn@doi [\aj]
  {10.1086/117748}, \href {http://adsabs.harvard.edu/abs/1995AJ....110.3073D}
  {110, 3073}

\bibitem[\protect\citeauthoryear{{Ertel} et~al.,}{{Ertel}
  et~al.}{2014}]{2014A&A...570A.128E}
{Ertel} S.,  et~al., 2014, \mn@doi [\aap] {10.1051/0004-6361/201424438}, \href
  {http://adsabs.harvard.edu/abs/2014A%26A...570A.128E} {570, A128}

\bibitem[\protect\citeauthoryear{{Ertel} et~al.,}{{Ertel}
  et~al.}{2016}]{2016arXiv160805731E}
{Ertel} S.,  et~al., 2016, preprint, \href
  {http://adsabs.harvard.edu/abs/2016arXiv160805731E} {} (\mn@eprint {arXiv}
  {1608.05731})

\bibitem[\protect\citeauthoryear{{Faramaz} et~al.,}{{Faramaz}
  et~al.}{2014}]{2014A&A...563A..72F}
{Faramaz} V.,  et~al., 2014, \mn@doi [\aap] {10.1051/0004-6361/201322469},
  \href {http://adsabs.harvard.edu/abs/2014A%26A...563A..72F} {563, A72}

\bibitem[\protect\citeauthoryear{{Faramaz}, {Beust}, {Augereau}, {Kalas}  \&
  {Graham}}{{Faramaz} et~al.}{2015}]{2015A&A...573A..87F}
{Faramaz} V.,  {Beust} H.,  {Augereau} J.-C.,  {Kalas} P.,   {Graham} J.~R.,
  2015, \mn@doi [\aap] {10.1051/0004-6361/201424691}, \href
  {http://adsabs.harvard.edu/abs/2015A%26A...573A..87F} {573, A87}

\bibitem[\protect\citeauthoryear{{Fern{\'a}ndez}}{{Fern{\'a}ndez}}{2005}]{2005ASSL..328.....F}
{Fern{\'a}ndez} J.~A.,  ed. 2005, {Comets - Nature, Dynamics, Origin and their
  Cosmological Relevance}  Astrophysics and Space Science Library Vol. 328,
  \mn@doi{10.1007/978-1-4020-3495-4.
}

\bibitem[\protect\citeauthoryear{{Fixsen} \& {Dwek}}{{Fixsen} \&
  {Dwek}}{2002}]{2002ApJ...578.1009F}
{Fixsen} D.~J.,  {Dwek} E.,  2002, \mn@doi [\apj] {10.1086/342658}, \href
  {http://adsabs.harvard.edu/abs/2002ApJ...578.1009F} {578, 1009}

\bibitem[\protect\citeauthoryear{{Ford} \& {Rasio}}{{Ford} \&
  {Rasio}}{2008}]{2008ApJ...686..621F}
{Ford} E.~B.,  {Rasio} F.~A.,  2008, \mn@doi [\apj] {10.1086/590926}, \href
  {http://adsabs.harvard.edu/abs/2008ApJ...686..621F} {686, 621}

\bibitem[\protect\citeauthoryear{{G{\'a}sp{\'a}r}, {Rieke}, {Su}, {Balog},
  {Trilling}, {Muzzerole}, {Apai}  \& {Kelly}}{{G{\'a}sp{\'a}r}
  et~al.}{2009}]{2009ApJ...697.1578G}
{G{\'a}sp{\'a}r} A.,  {Rieke} G.~H.,  {Su} K.~Y.~L.,  {Balog} Z.,  {Trilling}
  D.,  {Muzzerole} J.,  {Apai} D.,   {Kelly} B.~C.,  2009, \mn@doi [\apj]
  {10.1088/0004-637X/697/2/1578}, \href
  {http://adsabs.harvard.edu/abs/2009ApJ...697.1578G} {697, 1578}

\bibitem[\protect\citeauthoryear{{Gladman}, {Kavelaars}, {Petit}, {Morbidelli},
  {Holman}  \& {Loredo}}{{Gladman} et~al.}{2001}]{2001AJ....122.1051G}
{Gladman} B.,  {Kavelaars} J.~J.,  {Petit} J.-M.,  {Morbidelli} A.,  {Holman}
  M.~J.,   {Loredo} T.,  2001, \mn@doi [\aj] {10.1086/322080}, \href
  {http://adsabs.harvard.edu/abs/2001AJ....122.1051G} {122, 1051}

\bibitem[\protect\citeauthoryear{{Gomes}, {Levison}, {Tsiganis}  \&
  {Morbidelli}}{{Gomes} et~al.}{2005}]{2005Natur.435..466G}
{Gomes} R.,  {Levison} H.~F.,  {Tsiganis} K.,   {Morbidelli} A.,  2005, \mn@doi
  [\nat] {10.1038/nature03676}, \href
  {http://adsabs.harvard.edu/abs/2005Natur.435..466G} {435, 466}

\bibitem[\protect\citeauthoryear{{Hahn}, {Zook}, {Cooper}  \& {Sunkara}}{{Hahn}
  et~al.}{2002}]{2002Icar..158..360H}
{Hahn} J.~M.,  {Zook} H.~A.,  {Cooper} B.,   {Sunkara} B.,  2002, \mn@doi
  [\icarus] {10.1006/icar.2002.6881}, \href
  {http://adsabs.harvard.edu/abs/2002Icar..158..360H} {158, 360}

\bibitem[\protect\citeauthoryear{{Holland} et~al.,}{{Holland}
  et~al.}{1998}]{1998Natur.392..788H}
{Holland} W.~S.,  et~al., 1998, \mn@doi [\nat] {10.1038/33874}, \href
  {http://adsabs.harvard.edu/abs/1998Natur.392..788H} {392, 788}

\bibitem[\protect\citeauthoryear{{Ida} \& {Lin}}{{Ida} \&
  {Lin}}{2008}]{2008ApJ...673..487I}
{Ida} S.,  {Lin} D.~N.~C.,  2008, \mn@doi [\apj] {10.1086/523754}, \href
  {http://adsabs.harvard.edu/abs/2008ApJ...673..487I} {673, 487}

\bibitem[\protect\citeauthoryear{{Janson}, {Quanz}, {Carson}, {Thalmann},
  {Lafreni{\`e}re}  \& {Amara}}{{Janson} et~al.}{2015}]{2015A&A...574A.120J}
{Janson} M.,  {Quanz} S.~P.,  {Carson} J.~C.,  {Thalmann} C.,  {Lafreni{\`e}re}
  D.,   {Amara} A.,  2015, \mn@doi [\aap] {10.1051/0004-6361/201424944}, \href
  {http://adsabs.harvard.edu/abs/2015A%26A...574A.120J} {574, A120}

\bibitem[\protect\citeauthoryear{{Jenniskens}}{{Jenniskens}}{2006}]{2006mspc.book.....J}
{Jenniskens} P.,  2006, {Meteor Showers and their Parent Comets}

\bibitem[\protect\citeauthoryear{{Juri{\'c}} \& {Tremaine}}{{Juri{\'c}} \&
  {Tremaine}}{2008}]{2008ApJ...686..603J}
{Juri{\'c}} M.,  {Tremaine} S.,  2008, \mn@doi [\apj] {10.1086/590047}, \href
  {http://adsabs.harvard.edu/abs/2008ApJ...686..603J} {686, 603}

\bibitem[\protect\citeauthoryear{{Kelsall} et~al.,}{{Kelsall}
  et~al.}{1998}]{1998ApJ...508...44K}
{Kelsall} T.,  et~al., 1998, \mn@doi [\apj] {10.1086/306380}, \href
  {http://adsabs.harvard.edu/abs/1998ApJ...508...44K} {508, 44}

\bibitem[\protect\citeauthoryear{{Kimura} \& {Mann}}{{Kimura} \&
  {Mann}}{1998}]{1998EP&S...50..493K}
{Kimura} H.,  {Mann} I.,  1998, \mn@doi [Earth, Planets, and Space]
  {10.1186/BF03352140}, \href
  {http://adsabs.harvard.edu/abs/1998EP%26S...50..493K} {50, 493}

\bibitem[\protect\citeauthoryear{{Koerner}, {Sargent}  \& {Ostroff}}{{Koerner}
  et~al.}{2001}]{2001ApJ...560L.181K}
{Koerner} D.~W.,  {Sargent} A.~I.,   {Ostroff} N.~A.,  2001, \mn@doi [\apjl]
  {10.1086/324226}, \href {http://adsabs.harvard.edu/abs/2001ApJ...560L.181K}
  {560, L181}

\bibitem[\protect\citeauthoryear{{Lebreton} et~al.,}{{Lebreton}
  et~al.}{2013}]{2013A&A...555A.146L}
{Lebreton} J.,  et~al., 2013, \mn@doi [\aap] {10.1051/0004-6361/201321415},
  \href {http://adsabs.harvard.edu/abs/2013A%26A...555A.146L} {555, A146}

\bibitem[\protect\citeauthoryear{{Leinert}, {Roser}  \& {Buitrago}}{{Leinert}
  et~al.}{1983}]{1983A&A...118..345L}
{Leinert} C.,  {Roser} S.,   {Buitrago} J.,  1983, \aap, \href
  {http://adsabs.harvard.edu/abs/1983A%26A...118..345L} {118, 345}

\bibitem[\protect\citeauthoryear{{Levison} \& {Duncan}}{{Levison} \&
  {Duncan}}{1994}]{1994Icar..108...18L}
{Levison} H.~F.,  {Duncan} M.~J.,  1994, \mn@doi [\icarus]
  {10.1006/icar.1994.1039}, \href
  {http://cdsads.u-strasbg.fr/abs/1994Icar..108...18L} {108, 18}

\bibitem[\protect\citeauthoryear{{Lin} \& {Ida}}{{Lin} \&
  {Ida}}{1997}]{1997ApJ...477..781L}
{Lin} D.~N.~C.,  {Ida} S.,  1997, \apj, \href
  {http://adsabs.harvard.edu/abs/1997ApJ...477..781L} {477, 781}

\bibitem[\protect\citeauthoryear{{Lisse} et~al.,}{{Lisse}
  et~al.}{2012}]{2012ApJ...747...93L}
{Lisse} C.~M.,  et~al., 2012, \mn@doi [\apj] {10.1088/0004-637X/747/2/93},
  \href {http://adsabs.harvard.edu/abs/2012ApJ...747...93L} {747, 93}

\bibitem[\protect\citeauthoryear{{L{\"o}hne}, {Krivov}  \&
  {Rodmann}}{{L{\"o}hne} et~al.}{2008}]{2008ApJ...673.1123L}
{L{\"o}hne} T.,  {Krivov} A.~V.,   {Rodmann} J.,  2008, \mn@doi [\apj]
  {10.1086/524840}, \href {http://adsabs.harvard.edu/abs/2008ApJ...673.1123L}
  {673, 1123}

\bibitem[\protect\citeauthoryear{{Marboeuf}, {Bonsor}  \&
  {Augereau}}{{Marboeuf} et~al.}{2016}]{2016arXiv160403790M}
{Marboeuf} U.,  {Bonsor} A.,   {Augereau} J.-C.,  2016, \mn@doi [Planetary and
  Space Science] {10.1016/j.pss.2016.03.014}, \href
  {http://adsabs.harvard.edu/abs/2016arXiv160403790M} {}

\bibitem[\protect\citeauthoryear{{Marshall} et~al.,}{{Marshall}
  et~al.}{2016}]{2016arXiv160408286M}
{Marshall} J.~P.,  et~al., 2016, preprint, \href
  {http://adsabs.harvard.edu/abs/2016arXiv160408286M} {} (\mn@eprint {arXiv}
  {1604.08286})

\bibitem[\protect\citeauthoryear{{Masset} \& {Papaloizou}}{{Masset} \&
  {Papaloizou}}{2003}]{2003ApJ...588..494M}
{Masset} F.~S.,  {Papaloizou} J.~C.~B.,  2003, \mn@doi [\apj] {10.1086/373892},
  \href {http://adsabs.harvard.edu/abs/2003ApJ...588..494M} {588, 494}

\bibitem[\protect\citeauthoryear{{Mennesson} et~al.,}{{Mennesson}
  et~al.}{2014}]{2014ApJ...797..119M}
{Mennesson} B.,  et~al., 2014, \mn@doi [\apj] {10.1088/0004-637X/797/2/119},
  \href {http://adsabs.harvard.edu/abs/2014ApJ...797..119M} {797, 119}

\bibitem[\protect\citeauthoryear{{Moons} \& {Morbidelli}}{{Moons} \&
  {Morbidelli}}{1993}]{1993CeMDA..57...99M}
{Moons} M.,  {Morbidelli} A.,  1993, \mn@doi [Celestial Mechanics and Dynamical
  Astronomy] {10.1007/BF00692465}, \href
  {http://adsabs.harvard.edu/abs/1993CeMDA..57...99M} {57, 99}

\bibitem[\protect\citeauthoryear{{Moons} \& {Morbidelli}}{{Moons} \&
  {Morbidelli}}{1995}]{1995Icar..114...33M}
{Moons} M.,  {Morbidelli} A.,  1995, \mn@doi [\icarus]
  {10.1006/icar.1995.1041}, \href
  {http://adsabs.harvard.edu/abs/1995Icar..114...33M} {114, 33}

\bibitem[\protect\citeauthoryear{{Morbidelli} \& {Moons}}{{Morbidelli} \&
  {Moons}}{1995}]{1995Icar..115...60M}
{Morbidelli} A.,  {Moons} M.,  1995, \mn@doi [\icarus]
  {10.1006/icar.1995.1078}, \href
  {http://adsabs.harvard.edu/abs/1995Icar..115...60M} {115, 60}

\bibitem[\protect\citeauthoryear{{M{\"u}ller}, {L{\"o}hne}  \&
  {Krivov}}{{M{\"u}ller} et~al.}{2010}]{2010ApJ...708.1728M}
{M{\"u}ller} S.,  {L{\"o}hne} T.,   {Krivov} A.~V.,  2010, \mn@doi [\apj]
  {10.1088/0004-637X/708/2/1728}, \href
  {http://adsabs.harvard.edu/abs/2010ApJ...708.1728M} {708, 1728}

\bibitem[\protect\citeauthoryear{{Nesvorn{\'y}}, {Jenniskens}, {Levison},
  {Bottke}, {Vokrouhlick{\'y}}  \& {Gounelle}}{{Nesvorn{\'y}}
  et~al.}{2010}]{2010ApJ...713..816N}
{Nesvorn{\'y}} D.,  {Jenniskens} P.,  {Levison} H.~F.,  {Bottke} W.~F.,
  {Vokrouhlick{\'y}} D.,   {Gounelle} M.,  2010, \mn@doi [\apj]
  {10.1088/0004-637X/713/2/816}, \href
  {http://adsabs.harvard.edu/abs/2010ApJ...713..816N} {713, 816}

\bibitem[\protect\citeauthoryear{{Pearce} \& {Wyatt}}{{Pearce} \&
  {Wyatt}}{2014}]{2014MNRAS.443.2541P}
{Pearce} T.~D.,  {Wyatt} M.~C.,  2014, \mn@doi [\mnras]
  {10.1093/mnras/stu1302}, \href
  {http://adsabs.harvard.edu/abs/2014MNRAS.443.2541P} {443, 2541}

\bibitem[\protect\citeauthoryear{{Petit} et~al.,}{{Petit}
  et~al.}{2011}]{2011AJ....142..131P}
{Petit} J.-M.,  et~al., 2011, \mn@doi [\aj] {10.1088/0004-6256/142/4/131},
  \href {http://adsabs.harvard.edu/abs/2011AJ....142..131P} {142, 131}

\bibitem[\protect\citeauthoryear{{Pi{\'e}tu}, {di Folco}, {Guilloteau}, {Gueth}
   \& {Cox}}{{Pi{\'e}tu} et~al.}{2011}]{2011A&A...531L...2P}
{Pi{\'e}tu} V.,  {di Folco} E.,  {Guilloteau} S.,  {Gueth} F.,   {Cox} P.,
  2011, \mn@doi [\aap] {10.1051/0004-6361/201116796}, \href
  {http://adsabs.harvard.edu/abs/2011A%26A...531L...2P} {531, L2}

\bibitem[\protect\citeauthoryear{{Rasio} \& {Ford}}{{Rasio} \&
  {Ford}}{1996}]{1996Sci...274..954R}
{Rasio} F.~A.,  {Ford} E.~B.,  1996, \mn@doi [Science]
  {10.1126/science.274.5289.954}, \href
  {http://adsabs.harvard.edu/abs/1996Sci...274..954R} {274, 954}

\bibitem[\protect\citeauthoryear{{Raymond} \& {Bonsor}}{{Raymond} \&
  {Bonsor}}{2014}]{2014MNRAS.442L..18R}
{Raymond} S.~N.,  {Bonsor} A.,  2014, \mn@doi [\mnras] {10.1093/mnrasl/slu048},
  \href {http://adsabs.harvard.edu/abs/2014MNRAS.442L..18R} {442, L18}

\bibitem[\protect\citeauthoryear{{Raymond} et~al.,}{{Raymond}
  et~al.}{2011}]{2011A&A...530A..62R}
{Raymond} S.~N.,  et~al., 2011, \mn@doi [\aap] {10.1051/0004-6361/201116456},
  \href {http://adsabs.harvard.edu/abs/2011A%26A...530A..62R} {530, A62}

\bibitem[\protect\citeauthoryear{{Reach}, {Morris}, {Boulanger}  \&
  {Okumura}}{{Reach} et~al.}{2003}]{2003Icar..164..384R}
{Reach} W.~T.,  {Morris} P.,  {Boulanger} F.,   {Okumura} K.,  2003, \mn@doi
  [\icarus] {10.1016/S0019-1035(03)00133-7}, \href
  {http://adsabs.harvard.edu/abs/2003Icar..164..384R} {164, 384}

\bibitem[\protect\citeauthoryear{{Ricci}, {Carpenter}, {Fu}, {Hughes}, {Corder}
   \& {Isella}}{{Ricci} et~al.}{2015}]{2015ApJ...798..124R}
{Ricci} L.,  {Carpenter} J.~M.,  {Fu} B.,  {Hughes} A.~M.,  {Corder} S.,
  {Isella} A.,  2015, \mn@doi [\apj] {10.1088/0004-637X/798/2/124}, \href
  {http://adsabs.harvard.edu/abs/2015ApJ...798..124R} {798, 124}

\bibitem[\protect\citeauthoryear{{Rowan-Robinson} \& {May}}{{Rowan-Robinson} \&
  {May}}{2013}]{2013MNRAS.429.2894R}
{Rowan-Robinson} M.,  {May} B.,  2013, \mn@doi [\mnras] {10.1093/mnras/sts471},
  \href {http://adsabs.harvard.edu/abs/2013MNRAS.429.2894R} {429, 2894}

\bibitem[\protect\citeauthoryear{{Sibthorpe} et~al.,}{{Sibthorpe}
  et~al.}{2010}]{2010A&A...518L.130S}
{Sibthorpe} B.,  et~al., 2010, \mn@doi [\aap] {10.1051/0004-6361/201014574},
  \href {http://adsabs.harvard.edu/abs/2010A%26A...518L.130S} {518, L130}

\bibitem[\protect\citeauthoryear{{Su} et~al.,}{{Su}
  et~al.}{2005}]{2005ApJ...628..487S}
{Su} K.~Y.~L.,  et~al., 2005, \mn@doi [\apj] {10.1086/430819}, \href
  {http://adsabs.harvard.edu/abs/2005ApJ...628..487S} {628, 487}

\bibitem[\protect\citeauthoryear{{Su} et~al.,}{{Su}
  et~al.}{2006}]{2006ApJ...653..675S}
{Su} K.~Y.~L.,  et~al., 2006, \mn@doi [\apj] {10.1086/508649}, \href
  {http://adsabs.harvard.edu/abs/2006ApJ...653..675S} {653, 675}

\bibitem[\protect\citeauthoryear{{Su} et~al.,}{{Su}
  et~al.}{2013}]{2013ApJ...763..118S}
{Su} K.~Y.~L.,  et~al., 2013, \mn@doi [\apj] {10.1088/0004-637X/763/2/118},
  \href {http://adsabs.harvard.edu/abs/2013ApJ...763..118S} {763, 118}

\bibitem[\protect\citeauthoryear{{Vitense}, {Krivov}  \& {L{\"o}hne}}{{Vitense}
  et~al.}{2010}]{2010A&A...520A..32V}
{Vitense} C.,  {Krivov} A.~V.,   {L{\"o}hne} T.,  2010, \mn@doi [\aap]
  {10.1051/0004-6361/201014208}, \href
  {http://adsabs.harvard.edu/abs/2010A%26A...520A..32V} {520, A32}

\bibitem[\protect\citeauthoryear{{Ward}}{{Ward}}{1997}]{1997Icar..126..261W}
{Ward} W.~R.,  1997, \mn@doi [\icarus] {10.1006/icar.1996.5647}, \href
  {http://adsabs.harvard.edu/abs/1997Icar..126..261W} {126, 261}

\bibitem[\protect\citeauthoryear{{Weissman}}{{Weissman}}{1980}]{1980A&A....85..191W}
{Weissman} P.~R.,  1980, \aap, \href
  {http://adsabs.harvard.edu/abs/1980A%26A....85..191W} {85, 191}

\bibitem[\protect\citeauthoryear{{Wilner}, {Holman}, {Kuchner}  \&
  {Ho}}{{Wilner} et~al.}{2002}]{2002ApJ...569L.115W}
{Wilner} D.~J.,  {Holman} M.~J.,  {Kuchner} M.~J.,   {Ho} P.~T.~P.,  2002,
  \mn@doi [\apjl] {10.1086/340691}, \href
  {http://adsabs.harvard.edu/abs/2002ApJ...569L.115W} {569, L115}

\bibitem[\protect\citeauthoryear{{Wyatt}, {Dermott}, {Telesco}, {Fisher},
  {Grogan}, {Holmes}  \& {Pi{\~n}a}}{{Wyatt}
  et~al.}{1999}]{1999ApJ...527..918W}
{Wyatt} M.~C.,  {Dermott} S.~F.,  {Telesco} C.~M.,  {Fisher} R.~S.,  {Grogan}
  K.,  {Holmes} E.~K.,   {Pi{\~n}a} R.~K.,  1999, \mn@doi [\apj]
  {10.1086/308093}, \href {http://cdsads.u-strasbg.fr/abs/1999ApJ...527..918W}
  {527, 918}

\bibitem[\protect\citeauthoryear{{Wyatt}, {Smith}, {Greaves}, {Beichman},
  {Bryden}  \& {Lisse}}{{Wyatt} et~al.}{2007}]{2007ApJ...658..569W}
{Wyatt} M.~C.,  {Smith} R.,  {Greaves} J.~S.,  {Beichman} C.~A.,  {Bryden} G.,
   {Lisse} C.~M.,  2007, \mn@doi [\apj] {10.1086/510999}, \href
  {http://adsabs.harvard.edu/abs/2007ApJ...658..569W} {658, 569}

\bibitem[\protect\citeauthoryear{{Yoon}, {Peterson}, {Kurucz}  \&
  {Zagarello}}{{Yoon} et~al.}{2010}]{2010ApJ...708...71Y}
{Yoon} J.,  {Peterson} D.~M.,  {Kurucz} R.~L.,   {Zagarello} R.~J.,  2010,
  \mn@doi [\apj] {10.1088/0004-637X/708/1/71}, \href
  {http://adsabs.harvard.edu/abs/2010ApJ...708...71Y} {708, 71}

\makeatother
\end{thebibliography}


\bsp	
\label{lastpage}
\end{document}